\documentclass{SciPost}

\hypersetup{
    colorlinks,
    linkcolor={red!50!black},
    citecolor={blue!50!black},
    urlcolor={blue!80!black}
}

\usepackage[bitstream-charter]{mathdesign}
\DeclareSymbolFont{usualmathcal}{OMS}{cmsy}{m}{n}
\DeclareSymbolFontAlphabet{\mathcal}{usualmathcal}

\fancypagestyle{SPstyle}{
\fancyhf{}
\lhead{\colorbox{scipostblue}{\bf \color{white} ~SciPost Physics }}
\rhead{{\bf \color{scipostdeepblue} ~Submission }}

\fancyfoot[C]{\textbf{\thepage}}
}

\usepackage[utf8]{inputenc}
\usepackage{pifont}

\usepackage{amsmath}
\usepackage{mathtools}

\usepackage{array}

\usepackage{breqn} 
\makeatletter
\let\cat@comma@active\@empty
\makeatother

\usepackage{enumerate}
\usepackage[shortlabels]{enumitem}

\usepackage[caption=false]{subfig}

\usepackage{booktabs}
\usepackage{threeparttable}
\usepackage{tabularx}
\usepackage{multirow}
\usepackage{rotating}

\usepackage[dvipsnames,table]{xcolor}
\usepackage{graphicx}
\graphicspath{{figs/}} 

\usepackage{physics}                
\usepackage{csquotes}               

\usepackage[capitalize]{cleveref}
\usepackage{chngcntr} 

\usepackage{orcidlink}

\newcommand{\onlinecite}[1]{\citenum{#1}}

\renewcommand{\i}{\mathrm{i}}       

\newcommand{\Ham}{\mathcal{H}}      

\newcommand{\adjo}[1]{#1^\dagger}
\newcommand{\id}{\mathbb{1}}
\newcommand{\pdag}{\phantom{\dagger}}
\newcommand{\phasediagram}{$\alpha$-$z$ phase diagram}
\newcommand{\Phasediagram}{The $\alpha$-$z$ phase diagram}
\newcommand{\Phasediagrams}{The $\alpha$-$z$ phase diagrams}

\newcommand{\cmark}{\ding{51}} 
\newcommand{\xmark}{\ding{55}} 
\newcommand{\NA}{--} 

\usepackage{acronym}

\newacro{MI}{mutual information\acused{MI-ad}}
\newacroindefinite{MI}{an}{a}
\newacro{MI-ad}[MI]{mutual-information\acused{MI}}
\newacroindefinite{MI-ad}{an}{a}

\newacro{RMI}{Rényi mutual information\acused{RMI-ad}}
\newacroindefinite{RMI}{an}{a}
\newacro{RMI-ad}[RMI]{Rényi-mutual-information\acused{RMI}}
\newacroindefinite{RMI-ad}{an}{a}

\newacro{DPI}{data processing inequality\acused{DPI-ad}}
\newacro{DPI-ad}[DPI]{data-processing-inequality\acused{DPI}}

\newacro{1D}{one dimension\acused{1D-ad}}
\newacro{1D-ad}[1D]{one-dimensional\acused{1D}}

\newacro{CFT}{conformal field theory\acused{CFT-ad}}
\newacro{CFT-ad}[CFT]{conformal-field-theory\acused{CFT}}

\newacro{TFIM}{transverse-field Ising model}

\newacro{MPS}{matrix product state\acused{MPS-ad}}
\newacroplural{MPS}[MPS]{matrix product states\acused{MPS-ad}}
\newacro{MPS-ad}[MPS]{matrix-product-state\acused{MPS}}

\newacro{iMPS}{infinite matrix product state\acused{iMPS-ad}}
\newacroplural{iMPS}[iMPS]{infinite matrix product states\acused{iMPS-ad}}
\newacro{iMPS-ad}[iMPS]{infinite-matrix-product-state\acused{iMPS}}

\newacro{MPDO}{matrix product density operator}

\newcolumntype{Y}{>{\centering\arraybackslash}X}

\crefname{section}{Sec.}{Secs.}
\Crefname{section}{Section}{Sections}

\crefname{appendix}{App.}{Apps.}
\Crefname{appendix}{Appendix}{Appendices}

\begin{document}

\pagestyle{SPstyle}

\begin{center}{\Large \textbf{\color{scipostdeepblue}{%
Computational access to lattice and long-wavelength physics in quantum mutual information\\
}}}\end{center}

\begin{center}\textbf{%
Patrick M.\ Lenggenhager\,\orcidlink{0000-0001-6746-1387}\textsuperscript{1$\star$},
M.\ Michael Denner\,\orcidlink{0000-0002-1762-9687}\textsuperscript{2},
Doruk Efe Gökmen\,\orcidlink{0000-0002-5536-1941}\textsuperscript{3,4},
Maciej Koch-Janusz\,\orcidlink{0000-0002-2903-5202}\textsuperscript{5},
Titus Neupert\,\orcidlink{0000-0003-0604-041X}\textsuperscript{2}, and
Mark H.\ Fischer\,\orcidlink{0000-0003-0810-6064}\textsuperscript{2}
}\end{center}

\begin{center}
{\bf 1} Max Planck Institute for the Physics of Complex Systems, Dresden, Germany
\\
{\bf 2} Department of Physics, University of Zurich, Zürich, Switzerland
\\
{\bf 3} NSF-Simons National Institute for Theory and Mathematics in Biology, Chicago, USA
\\
{\bf 4} James Franck Institute and Department of Statistics, University of Chicago, Chicago, USA
\\
{\bf 5} Haiqu Inc., San Francisco, USA
\\[\baselineskip]
$\star$ \href{mailto:plengg@pks.mpg.de}{\small plengg@pks.mpg.de}
\end{center}

\section*{\color{scipostdeepblue}{Abstract}}
\textbf{\boldmath{%
Quantum mutual information is an important tool for characterizing correlations in quantum many-body systems, but its numerical evaluation is often prohibitively expensive. While some variants of Rényi Mutual Information (RMI) are computationally more tractable, it is not clear whether they correctly capture the long-wavelength physics or are dominated by UV effects, which is of key importance in lattice simulations. We analyze the relevance of lattice effects on the family of $\alpha$-$z$ Rényi mutual informations for ground states of models with conformal field theory descriptions. On the example of massless free fermions we identify distinct regions in the $\alpha$-$z$ plane, where RMI corrections due to the lattice are relevant or irrelevant. We further support these findings with MPS calculations on the transverse field Ising model (TFIM). Our results, accompanied by the open-source Julia package QMICalc.jl, provide guidance to using RMI in quantum-many body physics numerical computations.
}}

\vspace{\baselineskip}

\noindent\textcolor{white!90!black}{%
\fbox{\parbox{0.975\linewidth}{%
\textcolor{white!40!black}{\begin{tabular}{lr}%
  \begin{minipage}{0.6\textwidth}%
    {\small Copyright attribution to authors. \newline
    This work is a submission to SciPost Physics. \newline
    License information to appear upon publication. \newline
    Publication information to appear upon publication.}
  \end{minipage} & \begin{minipage}{0.4\textwidth}
    {\small Received Date \newline Accepted Date \newline Published Date}%
  \end{minipage}
\end{tabular}}
}}
}


\vspace{10pt}
\noindent\rule{\textwidth}{1pt}
\tableofcontents
\noindent\rule{\textwidth}{1pt}
\vspace{10pt}


\section{Introduction}\label{Sec:introduction}

Information-theoretic measures of correlations have proven useful for characterizing, understanding, and classifying quantum many-body systems~\cite{DeChiara2018,Frerot2023}, not only through ground-state but also through thermal and non-equilibrium properties.
In condensed-matter physics, for example, correlations play a central role in determining emergent properties, from quantum phase transitions to topological order and even dynamical behavior.
Moreover, the range and structure of these correlations may dictate the tools and models used to study a given system~\cite{Cirac2021}.
The length-scale dependence of correlations serves as a key diagnostic, providing insights into universality, criticality, and the interplay between short- and long-range effects.
Information theory offers a powerful framework for quantifying these correlations, adopting a perspective that extends beyond conventional two-point correlation functions.
(Quantum) Mutual information (MI)\acused{MI}\acused{MI-ad} encodes classical \emph{and} quantum correlations in a unified manner~\cite{Groisman2005}, capturing nonlocal and entanglement-driven effects that are essential for understanding criticality and phase structure.
This is in contrast to entanglement measures~\cite{Plbnio2007}, which focus on the quantum side exclusively.

\begin{figure}
    \centering
    \includegraphics{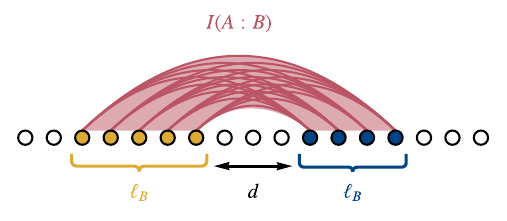}
    \caption{
        Partition of a chain (black circles) into two disjoint subsystems: $A$ (yellow disks) and $B$ (blue disks), with lengths $\ell_A$ and $\ell_B$, respectively, separated by a distance $d$, all measured in units of the lattice constant.
        The mutual information $I(A:B)$ encodes all correlations between the two subsystems, as illustrated by the red lines and shading connecting them.
    }
    \label{fig:partition_chain}
\end{figure}

The \ac{MI} $I(A:B)_\rho$ of a state described by the density matrix $\rho$ between two sets of degrees of freedom $A$ and $B$ quantifies the information shared between them.
By varying the partitioning of degrees of freedom into $A$, $B$, and the remainder of the system, one can map out the nature of correlations.
While $A$ and $B$ can represent any subsets of the degrees of freedom, they are often chosen as spatially disjoint subsystems.
\Cref{fig:partition_chain} illustrates such a partitioning in the example of a \ac{1D-ad} chain, where $I(A:B)_\rho$ measures the correlations between a subsystem $A$ of size $\ell_A$ and a subsystem $B$ of size $\ell_B$, separated by a distance $d$.

Mutual information exhibits several desirable physical characteristics, including area laws for various classes of states~\cite{Wolf2008,Kuwahara2020} and rigorous bounds on correlation functions~\cite{Wolf2008,Scalet2021}.
It has been used to probe for spontaneous symmetry breaking~\cite{Hamma2016} and many-body localization~\cite{DeTomasi2017} and has been shown to reflect universal properties at phase transitions~\cite{Alcaraz2013}.
In \ac{1D-ad} systems with their low-energy physics well described by \ac{CFT}, the \ac{MI} of ground states has been extensively studied~\cite{Calabrese2009,Calabrese2011}.
In addition to these analytical and physical insights, \ac{MI} has also found applications in numerical algorithms.
It has been shown to be a useful quantity to optimize for identifying relevant degrees of freedom and selecting optimal real-space renormalization group transformations~\cite{Tishby2000,KochJanusz2018,Lenggenhager2020,Gokmen2021,Gordon2021,Gokmen2024} and has been applied to overcome the entanglement barrier~\cite{Calabrese2005,Schuch2008,Lauchli2008} in time evolution of many-body systems utilizing the concept of the information lattice~\cite{Kvorning2022,Artiaco2024}.
These applications underscore the versatility of mutual information as a framework for quantifying the strength and nature of correlations across different settings.

The most common way to define \ac{MI} is in terms of the von Neumann entropy of the subsystem $A$, $S(A)_\rho=-\Tr(\rho_A\log\rho_A)$, where $\rho_A=\Tr_{\bar{A}}(\rho)$ is the reduced density matrix with the trace over the complement $\bar{A}$ of $A$.
In particular,
\begin{equation}
    I_{\mathrm{vN}}(A:B)_\rho = S(A)_\rho + S(B)_\rho - S(AB)_\rho.
    \label{eq:von-Neumann-MI:entropy}
\end{equation}
It is referred to as the \emph{von Neumann} \ac{MI}.
Despite its theoretical significance, computing it in practice is demanding, particularly for large systems.
This remains true even when using approximate methods, as the calculation requires diagonalizing reduced density matrices, a task that quickly becomes infeasible as system size grows.
To address this difficulty, alternative variants and definitions of \ac{MI} have been explored that avoid the computational bottleneck.
For instance, replacing the logarithm with power-law expressions gives rise to \emph{Rényi} MIs~\cite{Petz1985,Khatri2024,Audenaert2015}, which enable efficient variational approaches through tensor- and neural-network methods.
Despite these modifications, many essential physical properties are retained, including area-law behavior, bounds on correlation functions, and simplifications in the presence of conformal symmetry~\cite{Scalet2021,KudlerFlam2023,KudlerFlam2024}.

However, the differences in their behavior in practical settings, particularly in numerical simulations, remain less well understood.
In this work, we focus on the applicability and appropriateness of \ac{MI} as a tool to study and characterize phase transitions at quantum critical points.
While phase transitions are governed by long-wavelength (infrared; IR) physics, numerical simulations are often performed on lattice systems, where discretization effects due to the ultraviolet (UV) physics introduce deviations from continuum expectations.
Additionally, in many cases, the relevant states, such as ground states or thermal states, are not known exactly but are instead obtained through approximate numerical methods.
The extent to which different \ac{MI} variants are sensitive to such lattice effects and numerical approximations is therefore of great practical importance.
To address this question, we systematically investigate how different \ac{MI} variants respond to lattice effects, identifying qualitative and quantitative differences that can impact their reliability as correlation measures in numerical simulations.

\begin{figure}[p]
    \centering
    \subfloat{\label{fig:azRMI-phase-diagram_summary}}
    \subfloat{\label{fig:a-z-Renyi-MI-DPI}}
    \subfloat{\label{fig:I-vs-x_demo}}
    \includegraphics{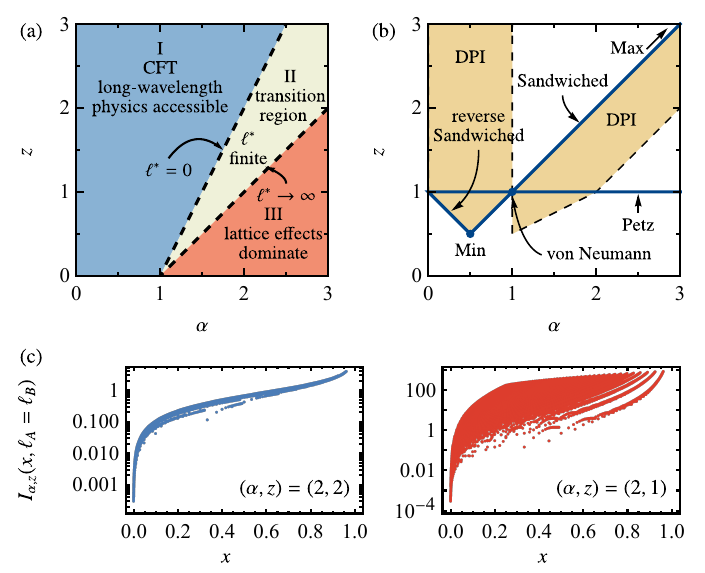}
    \caption{
    (a) Schematic \enquote{phase} diagram of the relevance of lattice effects in the $\alpha$-$z$ Rényi mutual information (RMI) for the ground state of models whose low-energy physics is described by a conformal field theory (CFT).
    Each point in the $\alpha$-$z$ plane corresponds to a distinct RMI.
    Variants from region I (blue) predominantly encode long-wavelength information, with lattice corrections that decay with the subsystem size $\ell$.
    In region III (red), on the other hand, lattice effects dominate, causing the mutual information to diverge with $\ell$ and rendering long-wavelength information inaccessible.
    The transition region II (light green) lies between these two, and each variant within it is associated with a finite length scale $\ell^*$ governing the accessibility of long-wavelength information.
    (b) Regions in the space of $\alpha$-$z$ RMIs, where the data processing inequality (DPI) [\cref{eq:DPI}] is proven or conjectured to be satisfied (shaded in yellow)~\cite{Audenaert2015}, along with well-known special cases (blue points and black arrows indicating points outside the displayed range) and one-parameter subfamilies (blue lines).
    (c) Two examples of $\alpha$-$z$ RMI $I_{\alpha,z}$: $(\alpha,z)=(2,2)$ (left) from region I, and $(\alpha,z)=(2,1)$ (right) from region III.
    Each data point corresponds to $I_{\alpha,z}(A:B)$ for the partition with $\ell_A=\ell_B=\ell$ and separation $d$, plotted as a function of the cross ratio $x$ [\cref{eq:cross-ratio}].
    The data is computed for the ground state of the free-fermion model given in \cref{eq:Hamiltonian-free-fermions}. In the left plot (region I) data points for different subsystem sizes collapse on a single curve determined by $x$ with only small corrections, while in the right one (region II) there is a strong dependence on the geometric quantities.
    }
    \label{fig:summary}
\end{figure}

In this work, we consider ground states of lattice models whose long-wavelength description is given by a \ac{CFT}.
This is typically the case at second-order phase transitions of many-body systems, where the conformal symmetry giving these field theories their name is closely related to the divergence of the correlation length at the phase transition.
The absence of an intrinsic length scale results in scale invariance.
In $1+1$ dimensions, conformal symmetry severely constrains the theory, which we exploit here to identify lattice effects.
For disjoint intervals ($d>0$), the \ac{MI} of a \ac{CFT-ad} ground state depends on the partition $\ell_A$, $d$, $\ell_B$ only through the \emph{cross-ratio}, defined as
\begin{equation}
    x = \frac{\ell_A\ell_B}{(\ell_A+d)(\ell_B+d)}.
    \label{eq:cross-ratio}
\end{equation}
Since conformal symmetry is broken on the lattice, we can attribute any explicit dependence on $\ell_A$, $d$, $\ell_B$ to lattice effects, distinguishing it from the long-wavelength physics.
With this insight, we can quantify and study those lattice effects.

We find that different variants of \ac{MI}, computed for the ground state of a \ac{CFT}, show qualitatively different responses to lattice effects.
In particular, we study a two-parameter family of \ac{RMI-ad} variants: the $\alpha$-$z$ \ac{RMI}.
We identify three \enquote{phases} of variants manifesting as regions in the $\alpha$-$z$ space of variants with distinct properties, as illustrated in \cref{fig:azRMI-phase-diagram_summary}.
Region I contains \ac{MI-ad} variants that predominantly capture the long-wavelength physics with lattice effects showing up only as irrelevant corrections vanishing for sufficiently large subsystems.
This makes the universal long-wavelength physics, in principle, accessible at any scale.

In contrast, variants in region III are dominated by non-universal short-wavelength physics, rendering long-wavelength information completely inaccessible.
Additionally, these variants violate important information-theoretic properties expected of faithful measures of correlation, as well as physical properties such as area laws.
Finally, there is an extended transition region (II) interpolating between regions I and III, where access to the long-wavelength physics depends on the subsystem size and partitioning.
Understanding which variants are sensitive to short- vs.\ long-wavelength physics, i.e., where lattice contributions are relevant vs.\ irrelevant, is crucial in choosing the appropriate one for a specific application, especially since artificial lattice effects can also arise as a consequence of approximations.

More specifically, we demonstrate this behavior in two concrete models and support our claim of universality of the regions with arguments based purely on \ac{CFT}.
First, we consider massless free fermions, which permit a numerically efficient and exact computation of the \ac{MI-ad} based on the ground-state correlation matrix known analytically.
Second, we study the \ac{TFIM}~\cite{Sachdev2011}, a noninteracting model, which, while exactly solvable, does not permit efficient computation of \ac{MI}.
We thus utilize \acp{MPS}~\cite{Schollwock2011} to compute the \ac{MI}.
Since \acp{MPS} can accurately describe the ground state of a broad class of systems, this demonstrates that our observations likely have practical implications well beyond integrable models.
Additionally, we uncover a dependence of the convergence properties of \ac{MI} with bond dimension on the variants, mirroring their sensitivity to lattice effects.
In this context, we have developed an open-source Julia package \textsc{QMICalc}~\cite{QMICalc} implementing the computation of \ac{MI-ad} variants for any given \ac{iMPS}.

The remainder of the paper is structured as follows.
In \cref{Sec:MI-variants}, we review common variants of \ac{MI} and summarize their key properties in \cref{tab:mi-properties}.
\Cref{Sec:lattice-effects} provides a qualitative overview of our main results, summarized in \cref{fig:summary}, and arguments for their generality.
In \cref{Sec:free-fermions,Sec:TFIM}, we study massless free fermions and the \ac{TFIM}, respectively, obtaining the results summarized before.
Finally, we conclude in \cref{Sec:conclusions} with an outlook.
Technical and numerical details are deferred to the appendices.

\section{Variants of mutual information}\label{Sec:MI-variants}

In this section, we give an overview of different variants of quantum mutual information.
We define several well-known families of \ac{MI-ad} variants and illustrate those within the subspace of variants that we focus on in \cref{fig:a-z-Renyi-MI-DPI}.
Additionally, we summarize and compare the most important properties of several variants in \cref{tab:mi-properties}.

Quantum mutual information can be axiomatically defined as a measure of correlations.
Given a state described by the density matrix $\rho$, mutual information $I(A:B)_\rho$ quantifies correlations between two sets of degrees of freedom, $A$ and $B$.
The defining axioms include symmetry in $A$ and $B$, unitary invariance, and continuity in $\rho$, as well as the following properties:
(1) \ac{MI} vanishes if and only if $\rho_{AB}=\rho_A\otimes\rho_B$, ensuring that independent systems are correctly identified with the uncorrelated ones, and
(2) \ac{MI} does not increase under local operations, in other words under the action of quantum channels $\mathcal{M}$ and $\mathcal{N}$ acting independently on $A$ and $B$, respectively~\cite{Khatri2024}.
The latter property is known as the \emph{\ac{DPI}}
\begin{equation}
    I(A:B)_\rho \geq I(A:B)_{(\mathcal{M}\otimes\mathcal{N})(\rho)}.
    \label{eq:DPI}
\end{equation}
Another important property of a correlation measure, nonnegativity [$I(A:B)_\rho \geq 0$], follows from \ac{DPI} by choosing the quantum channels to be partial traces, $\rho_{AB}\mapsto\rho_A\otimes\rho_B$, which renders $A$ and $B$ independent in the output state.

\begin{sidewaystable}
\footnotesize
\centering
\begin{threeparttable}
\caption{
    Properties of variants of quantum mutual information, in particular Rényi mutual information (RMI). If the property is satisfied (for any $\alpha$), this is indicated by \enquote{\cmark}, if it is generally violated by \enquote{\xmark}, and if it holds for a range of $\alpha$, that range is given instead. A question mark indicates that to the best of our knowledge, it is unknown. If no other reference is given, information-theoretic properties (rows 1 to 3) can be found in Ref.~\onlinecite{Khatri2024}; bounds (rows 4 to 6) often follow from bounds on related quantities, see table footnotes.
    The data processing inequality is defined in \cref{eq:DPI}, $\alpha$-monotonicity refers to monotonic growth with $\alpha$, and special cases refer to limits of $\alpha$ corresponding to the well-known variants.
}
\label{tab:mi-properties}
\begin{tabularx}{\textwidth}{@{\extracolsep{\fill}} p{2.75cm}YYYYYYY}
    \toprule
    \textbf{RMI variant} & \multirow{2}{2cm}{\centering entropy-based} &  \multicolumn{4}{c}{subfamilies and special cases of $\alpha$-$z$ RMI $I_{\alpha,z}$} & measured & geometric \\
    \textbf{Property} & & $I_{1,1}=I_{\mathrm{vN}}$ & $I_{\alpha,1}$ & $I_{\alpha,\alpha}$ & $I_{\infty,\infty}$ & $I_\alpha^{\mathbb{M}}$ & $I_\alpha^{\mathrm{G}}$ \\
    \midrule
    Data processing / nonnegativity & $(1,2)$\tnote{a}~\cite{Scalet2021} & \cmark & $[0,1)\cup(1,2]$ & $[\frac{1}{2},1)\cup(1,\infty)$\tnote{a} & \cmark & \cmark~\cite{Scalet2021} & $(0,1)\cup(1,2]$ \\
    $\alpha$-Monotonicity & \xmark~\cite{Kormos2017,Scalet2021} & \NA & \cmark & \cmark & \NA & \cmark~\cite{Rippchen2024} & \cmark \\
    Special cases of $\alpha$ & $1$~\cite{Calabrese2004} & \NA & $1$ & $\frac{1}{2},1,\infty$~\cite{MuellerLennert2013} & \NA & $\frac{1}{2},\infty$~\cite{Berta2017} & $1,\infty$\tnote{b} \\
    Thermal area laws & ? & \cmark~\cite{Wolf2008} & $(0,1)\cup(1,2]$\tnote{c} & \cmark\tnote{d} & \cmark~\cite{Scalet2021} & \cmark\tnote{d}~\cite{Scalet2021} & \cmark\tnote{d} \\
    PEPDO area laws & ? & \cmark~\cite{Wolf2008} & $(0,1)\cup(1,\frac{3}{2}]$~\cite{Scalet2021} & $[\frac{1}{2},1)\cup(1,2]$~\cite{Scalet2021} & \xmark~\cite{Scalet2021} & $(0,1)\cup(1,2]$~\cite{Scalet2021} & ? \\
    Correlation bounds & ? & \cmark~\cite{Wolf2008} & \cmark~\cite{KudlerFlam2023} & $[\frac{1}{2},1)\cup(1,\infty)$\tnote{e} & \cmark\tnote{d}~\cite{Scalet2021} & \cmark~\cite{Scalet2021} & $(0,1)\cup(1,2]$\tnote{c} \\
    CFT predictions & \cmark~\cite{Calabrese2009,Calabrese2011} & \cmark~\cite{Calabrese2009,Calabrese2011} & \cmark~\cite{KudlerFlam2023} & \cmark~\cite{KudlerFlam2024} & ? & ? & ? \\
    \bottomrule
\end{tabularx}
\begin{tablenotes}
    \item[a] Known to be violated otherwise.
    \item[b] For $\alpha\to 1$, $I_\alpha^{\mathrm{G}}$ converges to the Belavkin–Staszewski mutual information~\cite{Khatri2024}, which is inequivalent to $I_{1,1}$.
    \item[c] Follows from $\forall\alpha\in(0,1)\cup(1,2]$: $I_{\alpha,1}\leq I_\alpha^{\mathrm{G}}$
    \item[d] Follows from $\forall\alpha\in(0,1)\cup(1,\infty)$: $I_\alpha^{\mathbb{M}},I_{\alpha,\alpha},I_\alpha^{\mathrm{G}} \leq I_{\infty,\infty}$.
    \item[e] Follows from $\forall\alpha\in(\frac{1}{2},1)\cup(1,\infty)$: $I_\alpha^{\mathbb{M}}\leq I_{\alpha,\alpha}$.
\end{tablenotes}
\end{threeparttable}
\end{sidewaystable}

Among the quantities that satisfy these axioms, the most widely used is the von Neumann \ac{MI}, defined in \cref{eq:von-Neumann-MI:entropy}, which naturally extends the classical (Shannon) \ac{MI} to quantum systems.
The role of $I_{vN}(A:B)_\rho$ (which we shall henceforth denote by $I_{1,1}(A:B)_\rho$) as a correlation measure is further supported by its connection to two-point correlation functions.
Specifically, it provides an upper bound on the connected two-point correlation function for arbitrary operators $O_A$ and $O_B$ supported on $A$ and $B$, respectively~\cite{Wolf2008}:
\begin{equation}
    \frac{\left(\expval{O_AO_B}-\expval{O_A}\expval{O_B}\right)^2}{2\norm{O_A}^2_\infty\norm{O_B}^2_\infty} \leq I_{1,1}(A:B),
\end{equation}
where $\norm{\cdot}_\infty$ denotes the operator norm.

To circumvent the computational challenges posed by the logarithm of the density matrix, the von Neumann entropy is often replaced by the Rényi entropy~\cite{Renyi1961}, defined as
\begin{equation}
    S_\alpha(A)_\rho=\frac{1}{1-\alpha}\log\Tr\rho_A^\alpha,
    \label{eq:Renyi-entropy}
\end{equation}
which allows for more efficient evaluation of powers of the reduced density matrix.
Although a definition analogous to \cref{eq:von-Neumann-MI:entropy}, with $S$ replaced by $S_\alpha$, converges to the von Neumann \ac{MI} as $\alpha\to 1$~\cite{Calabrese2004}, it does not satisfy the \ac{DPI} for general $\alpha$ and even violates nonnegativity~\cite{Linden2013,Kormos2017,Scalet2021}.
This implies that these \emph{entropy-based} \acp{RMI} are not faithful measures of correlations.
Despite this limitation, they have been widely used in condensed matter physics~\cite{Wolf2008,Calabrese2009,Calabrese2011,Alcaraz2013}.
One particular application is the analytical continuation of the Rényi parameter from integer values $\alpha\in\mathbb{N}_{\geq2}$ to compute the limit $\alpha\to 1$.
However, such analytical continuation is often challenging, making alternative approaches more fruitful, especially those that facilitate numerical studies and allow for physical interpretations even at $\alpha\neq 1$.

To overcome the limitations of entropy-based definitions, \ac{MI} can be defined in terms of measures of state distinguishability, rather than relying on entropies.
In this approach, \ac{MI} is expressed in terms of a \emph{divergence} or \emph{relative entropy} $D(\rho\|\sigma)$, which quantifies the difference between two positive semi-definite operators, $\rho$ and $\sigma$.
Defining \ac{MI} as
\begin{equation}
    I(A:B)_\rho = D(\rho_{AB}\|\rho_A\otimes\rho_B),
\end{equation}
ensures that it satisfies the required axioms.
Different choices of divergence then give rise to distinct \ac{MI-ad} variants.
For instance, the von Neumann mutual information, \cref{eq:von-Neumann-MI:entropy}, is obtained from Umegaki's relative entropy~\cite{Umegaki1962}
\begin{equation}
    D_{\mathrm{U}}(\rho\|\sigma) = \Tr(\rho\log\rho-\rho\log\sigma),
    \label{eq:von-Neumann-divergence}
\end{equation}
which serves as the quantum generalization of the Kullback-Leibler divergence~\cite{Kullback1951}.

Similarly, quantum generalizations of the classical Rényi divergence provide a foundation for defining quantum \ac{RMI} $I_\alpha(A:B)_\rho$.
For example, a direct extension to density matrices is given by the \emph{Petz}-Rényi divergence~\cite{Petz1985,Khatri2024}
\begin{equation}
    D_{\alpha}(\rho\|\sigma) = \frac{1}{\alpha-1}\log\Tr\left(\rho^\alpha\sigma^{1-\alpha}\right).
    \label{eq:Petz-Renyi-divergence}
\end{equation}
However, this divergence and the resulting \ac{MI} satisfy \ac{DPI} only for $\alpha\in(1,2)$~\cite{Scalet2021}.
This limitation has motivated the exploration of alternative variants with the same classical limit.
In particular, exploiting the freedom in operator ordering inside the trace yields alternative divergences and thus additional \ac{RMI-ad} variants.

A broad range of operator orderings, along with several well-known limiting cases such as the minimal and maximal \ac{MI}, are unified in the two-parameter family of $\alpha$-$z$ \ac{RMI}~\cite{Audenaert2015}, defined as
\begin{subequations}
    \label{eq:a-z-Renyi-MI}
    \begin{equation}
        I_{\alpha,z}(A:B)_\rho = D_{\alpha,z}(\rho_{AB}\|\rho_A\otimes\rho_B)
        \label{eq:a-z-Renyi-MI:MI}
    \end{equation}
    with
    \begin{equation}
        D_{\alpha,z}(\rho\|\sigma) = \frac{1}{\alpha-1}\log\Tr\left(\sigma^{\frac{1-\alpha}{2z}}\rho^{\frac{\alpha}{z}}\sigma^{\frac{1-\alpha}{2z}}\right)^z.
        \label{eq:a-z-Renyi-MI:divergence}
    \end{equation}
\end{subequations}
Rewriting the argument of the trace in \cref{eq:a-z-Renyi-MI:divergence} as $(\rho^{\alpha/z}\sigma^{(1-\alpha)/z})^z$ shows that $z$ determines the ordering of operators inside the trace.
Notably, the choice $z=1$ recovers the Petz \ac{RMI}, \cref{eq:Petz-Renyi-divergence}, and the limit $\alpha=z\to 1$ corresponds to the von Neumann mutual information, \cref{eq:von-Neumann-MI:entropy,eq:von-Neumann-divergence}.
The \ac{DPI} is rigorously proven or conjectured to hold within specific regions of the $\alpha$-$z$ parameter space~\cite{Audenaert2015}, as illustrated in \cref{fig:a-z-Renyi-MI-DPI}.

Two additional variants, not part of the family in \cref{eq:a-z-Renyi-MI}, are worth mentioning.
The geometric \ac{RMI} is obtained from the divergence~\cite{Khatri2024}
\begin{equation}
    D_\alpha^{\mathrm{G}}(\rho\|\sigma) = \frac{1}{\alpha-1}\log\Tr\left[\sigma\left(\sigma^{-1/2}\rho\sigma^{-1/2}\right)^\alpha\right],
    \label{eq:geometric-RMI}
\end{equation}
which corresponds to an operator ordering not covered by \cref{eq:a-z-Renyi-MI:divergence} and converges to the Belavkin–Staszewski mutual information~\cite{Khatri2024} for $\alpha\to 1$ and not to the von Neumann \ac{MI}.
In contrast, the \emph{measured} \ac{RMI}~\cite{Scalet2021} adopts a fundamentally different approach to generalizing classical \ac{RMI}.
Rather than directly generalizing the classical expression, it is applied to the probability distributions obtained from a POVM measurement~\cite{Berta2017}:
\begin{equation}
    D_\alpha^{\mathbb{M}}(\rho\|\sigma) = \sup_{(\chi,M)}D_\alpha(P_{\rho,M}\|P_{\sigma,M}),
    \label{eq:measured-RMI}
\end{equation}
where the supremum is over finite sets $\chi$ of measurement outcomes and positive operator-valued measures (POVMs) $M$ on $\chi$; $P_{\cdot,M}$ denotes the post-measurement probability distribution, i.e., $P_{\rho,M}(x) = \Tr(M(x)\rho)$ for $x\in\chi$. 
The measured \ac{RMI} can also be used to construct variational estimators based on neural network critic functions~\cite{Goldfeld2024} via quantum generalizations of the Donsker-Varadhan formula.

While several \ac{MI-ad} variants have been explored with respect to physical properties such as area laws, correlation bounds, and simplifications for $(1+1)$D \acp{CFT} (see \cref{tab:mi-properties}), a comparison of variants in the context of the following two key questions is still lacking.
(1) What information about the physical system is encoded in the dependence of \ac{MI} on different spatial partitions in a ground state?
(2) How robust is this information with respect to short-wavelength properties of the system and deviations of the state from the true ground state?
As a first step toward answering these broader questions, we focus on the influence of lattice effects on the behavior of different \ac{MI-ad} variants in a system whose low-energy physics is captured by a \ac{CFT}.
Since the $\alpha$-$z$ \ac{RMI} includes most well-known variants within a convenient two-parameter family, we perform our systematic analysis in this space, parameterized by $\alpha$ and $z$ (see \cref{fig:a-z-Renyi-MI-DPI}).

\section{Lattice effects on mutual information}\label{Sec:lattice-effects}

In this section, we summarize our results on how different \ac{RMI-ad} variants respond qualitatively to lattice effects.
In particular, we study the relevance of lattice contributions to the $\alpha$-$z$ \ac{RMI} [\cref{eq:a-z-Renyi-MI}] in the ground state of $(1+1)$D \acp{CFT}.
While our quantitative results are based on massless free fermions (\cref{Sec:free-fermions}), we conjecture that the observations summarized in \cref{fig:azRMI-phase-diagram_summary} are universal within the \ac{CFT-ad} setting and thus model independent.
In \cref{Sec:universality-of-regions}, we justify this conjecture using arguments based on conformal symmetry and the mathematical structure of \cref{eq:a-z-Renyi-MI}, and later support it by verifying our observations in the \ac{TFIM} (\cref{Sec:TFIM}).

\subsection{Summary of results}\label{Sec:summary-of-results}

Our main result is that the space of $\alpha$-$z$ \ac{RMI-ad} variants, parameterized by $\alpha$ and $z$, splits into three regions (\cref{fig:azRMI-phase-diagram_summary}), distinguished by relevance or irrelevance of lattice contributions to the \ac{RMI} computed for a ground state of a model with a \ac{CFT} as low-energy description.
As a consequence, a variant's accessibility to long-wavelength information is determined by the region it lies in.
Region I contains variants dominated by the universal long-wavelength (\ac{CFT-ad}) behavior, with lattice effects appearing as irrelevant corrections, namely corrections that vanish in the limit of large subsystems.
This makes long-wavelength information accessible, either directly or through extrapolation.
Variants from region III, on the other hand, are dominated by lattice contributions, which diverge in the large-subsystem limit, rendering the universal long-wavelength information inaccessible and suggesting violation of area laws.

Interpolating between these two regions is a transition region (II), where the relevance of lattice contributions, and hence access to the long-wavelength information, depends on the partition.
The interpolation is governed by a characteristic length scale $\ell^*$, assigned \emph{to each \ac{MI-ad} variant}, and dependent on the partition.
This length scale, which we call the \emph{resolution length scale}, quantifies the minimal subsystem size required for a variant to resolve the long-wavelength physics.
Consistent with the phenomenology in the other regions, $\ell^*$ decreases toward region I, vanishing at the boundary, and diverges somewhere in region II, with the precise location depending on the partition.
This divergence is accompanied by a logarithmic dependence of the \ac{MI} on $\ell$.
At the boundary with region III, $\ell^*$ diverges for all possible choices of partition.

To make these observations more precise, we exploit the fact that conformal symmetry restricts the dependence of \ac{MI} on the partition parameters $\ell_A$, $d$, and $\ell_B$.
This enables us to separate lattice effects from the long-wavelength (\ac{CFT-ad}) contribution.
For disjoint intervals ($d>0$), the \ac{MI} of a system with conformal symmetry depends on the partition only through the cross-ratio, defined in \cref{eq:cross-ratio}.
This implies that in a \ac{CFT}
\begin{equation}
    I^{\mathrm{CFT}}(A:B) = I^{\mathrm{CFT}}(\ell_A,d,\ell_B) = I^{\mathrm{CFT}}(x).
    \label{eq:MI-CFT}
\end{equation}

On a lattice, however, conformal symmetry, particularly scale invariance, is broken.
Thus, if $\ell_A,d,\ell_B\gg 1$ does not hold (with distances measured in units of the lattice constant), the relation $I(A:B) = I(x)$ is no longer true.
Consequently, $I(A:B)$ deviates from the long-wavelength value $I^{\mathrm{CFT}}(A:B)$ and we attribute these deviations to lattice effects.
This motivates decomposing the mutual information of a lattice model into a universal long-wavelength (\ac{CFT-ad}) contribution $I^{\mathrm{CFT}}(x)$, which depends only on the cross-ratio $x$, and a lattice contribution $\Delta I(x,\ell_A,\ell_B)$, which explicitly depends on the partition,
\begin{equation}
    I^{\mathrm{lat}}(A:B) = I^{\mathrm{CFT}}(x) + \Delta I(x,\ell_A,\ell_B).
    \label{eq:MI-decomposition}
\end{equation}

\Cref{fig:I-vs-x_demo} illustrates how an explicit dependence on the partition parameters, beyond $x$, reveals qualitatively different behaviors for two representative variants.
If the subsystems are constrained to equal size ($\ell_A=\ell_B=\ell$) and their separation $d$ is determined by the cross-ratio $x$ [\cref{eq:cross-ratio}], the data in a plot of $I_{\alpha,z}$ versus $x$ is expected to collapse onto a single curve when lattice contributions vanish, i.e., $\Delta I=0$.
This expectation is indeed approximately satisfied for variants in region I, with only small corrections (\cref{fig:I-vs-x_demo}, top panel).
In contrast, this behavior breaks down completely for variants in region III (\cref{fig:I-vs-x_demo}, bottom panel).

\subsection{Universality of regions}\label{Sec:universality-of-regions}

We now provide arguments to justify our conjecture that the regions depicted in \cref{fig:azRMI-phase-diagram_summary}, including the black dashed boundaries, as well as the qualitative behavior of \ac{MI-ad} variants within each region are model independent and apply to any \ac{1D} lattice model with a \ac{CFT} low-energy description.

In the special case of adjacent intervals ($d=0$), corresponding to $x=1$, conformal symmetry alone determines the leading term of the \ac{MI} completely~\cite{KudlerFlam2024}, independent of the model (as before, we set $\ell_A=\ell_B=\ell$ for simplicity):
\begin{subequations}\label{eq:azRMI_d=0_CFT}
    \begin{align}
        I_{\alpha,z}(x=1,\ell) &= I_{\alpha,z}^{\mathrm{CFT}}(1,\ell) + \Delta I_{\alpha,z}(1,\ell)\\
        \intertext{with}
        I_{\alpha,z}^{\mathrm{CFT}}(1,\ell) &= c\frac{z+1-\alpha}{3(z+2-2\alpha)}\log\frac{\ell}{2\epsilon},\\
        \Delta I_{\alpha,z}(1,\ell) &= \frac{1}{\alpha-1}\log C_{\alpha,z}(\ell).
    \end{align}
\end{subequations}
Here, $c$ is the central charge of the \ac{CFT}, $C_{\alpha,z}$ is a universal operator product expansion coefficient, and $\epsilon$ a regularizer which can be identified with the lattice constant (chosen to be $1$ here).
Note that for $d=0$, in contrast to $d>0$, $I_{\alpha,z}^{\mathrm{CFT}}$ explicitly depends on $\ell$.

The prefactor $(z+1-\alpha)/(z+2-2\alpha)$ in $I_{\alpha,z}^{\mathrm{CFT}}$ exhibits a sign structure that precisely matches the regions in \cref{fig:azRMI-phase-diagram_summary}.
The prefactor, and thus $I_{\alpha,z}^{\mathrm{CFT}}$ (assuming $\ell>2$), is positive in regions I and III but negative in region II.
Since the full mutual information $I_{\alpha,z}$ is nonnegative (at least in the region with \ac{DPI} satisfied, see \cref{fig:a-z-Renyi-MI-DPI}), a negative $I_{\alpha,z}^{\mathrm{CFT}}$ that grows logarithmically in magnitude implies that the remaining term, $\Delta I_{\alpha,z}$, must also grow at least logarithmically.
This strongly suggests that $\Delta I_{\alpha,z}$ is relevant for at least $x=1$.
Along the boundary to region III, this effect is accentuated by the vanishing prefactor, making $\Delta I_{\alpha,z}(1,\ell)$ the sole remaining contribution.
Conversely, the prefactor diverges along the boundary to region I, rendering $I_{\alpha,z}^{\mathrm{CFT}}$ dominant and $\Delta I_{\alpha,z}$ irrelevant.

In contrast, the behavior in region III, defined by $z<\alpha-1$, can be understood purely from the definition of the $\alpha$-$z$ \ac{RMI} [\cref{eq:a-z-Renyi-MI}], without requiring conformal symmetry.
The reduced density matrices $\rho_{A,B}$ for the separate subsystems appear with an overall exponent of $\frac{1-\alpha}{z}$.
In region III, where $\frac{\alpha-1}{z}>1$, this exponent is less than $-1$, implying that small eigenvalues of $\rho_{A,B}$ dominate.
Presumably, lattice effects are encoded in such small eigenvalues~\cite{Calabrese2008}, which explains the marked sensitivity of \ac{MI-ad} variants in this region to such effects.
In particular, as $\ell\to\infty$, the eigenvalues are expected to vanish, causing the observed divergence of \ac{MI}.

\subsection{Application: choice of correlation measure}\label{Sec:choice-of-variant}

In combination with the known regions where \ac{DPI} is satisfied (\cref{fig:a-z-Renyi-MI-DPI}), our results offer practical guidance for selecting \iac{RMI-ad} variant or subfamily appropriate to a given application.
Comparing with \cref{fig:azRMI-phase-diagram_summary}, we observe that, while regions I--III do not perfectly align with the regions where \ac{DPI} is satisfied, there are significant intersections between \ac{DPI} and regions I and II.

Most of region I with $\alpha\leq 1$ satisfies \ac{DPI}, whereas only a small portion intersects with \ac{DPI} for $\alpha>1$.
A variant from this intersection can access the universal long-wavelength information governed by the \ac{CFT}.
In contrast, variants from the intersection of region II with \ac{DPI} are highly sensitive to the geometry of the partition, which determines access to long-wavelength physics.
This sensitivity could be exploited in applications targeting information at specific length scales.
Finally, all variants in region III violate \ac{DPI}, reinforcing that their sensitivity to short-wavelength physics and divergence as $\ell\to\infty$ preclude them from serving as reliable correlation measures.

\section{Massless free fermions}\label{Sec:free-fermions}

We now turn to the study of a lattice model whose long-wavelength physics is captured by the simplest \ac{CFT}, namely massless free fermions in \ac{1D}.
The quantitative numerical results obtained for this model provide the basis for the conclusions described in \cref{Sec:summary-of-results}.

\subsection{Model}\label{Sec:free-fermions:model}

We consider massless free fermions with annihilation and creation operators satisfying the anticommutation relation $\acomm*{c^{\pdag}_i}{\adjo{c_j}}=\delta_{ij}$.
The Hamiltonian is given by:
\begin{equation}
    \Ham = -\frac{\i}{2}\sum_j\left(\adjo{c_j}c^\pdag_{j+1}-\adjo{c_{j+1}}c^\pdag_j\right),
    \label{eq:Hamiltonian-free-fermions}
\end{equation}
where the chosen gauge emphasizes the associated low-energy field theory of two chiral fermions.
Since the Hamiltonian is quadratic in the fermionic operators, its eigenstates are Gaussian states fully characterized by two-point correlation functions $C_{ij}=\expval*{\adjo{c_i}c^\pdag_j}$~\cite{Peschel2003}.

This allows us to express the $\alpha$-$z$ \ac{RMI} in terms of $(\ell_A+\ell_B)\times(\ell_A+\ell_B)$ correlation matrices rather than the corresponding $2^{\ell_A+\ell_B}\times 2^{\ell_A+\ell_B}$ density matrices.
For the state $\rho_{AB}$, the correlation matrix is
\begin{equation}
    C = \mqty(C_{AA}&C_{AB}\\C_{BA}&C_{BB}),
    \label{eq:a-z-Renyi-MI_free-fermions:C}
\end{equation}
and for $\rho_A\otimes\rho_B$, it is
\begin{equation}
    C' = \mqty(C_{AA}&0\\0&C_{BB}).
    \label{eq:a-z-Renyi-MI_free-fermions:Cprime}
\end{equation}
The $\alpha$-$z$ \ac{RMI} can thus be written~\cite{Banchi2014,Casini2018,KudlerFlam2024} in the form
\begin{dmath}
    I_{\alpha,z}(A:B)_\rho = -\frac{\alpha}{1-\alpha}\Tr\log(\id-C) - \Tr\log(\id-C') - \frac{1}{1-\alpha}\Tr\log\left\{\id + \left[\left(\frac{C'}{\id-C'}\right)^{\frac{1-\alpha}{2z}}\left(\frac{C}{\id-C}\right)^{\frac{\alpha}{z}}\left(\frac{C'}{\id-C'}\right)^{\frac{1-\alpha}{2z}}\right]^z\right\}.
    \label{eq:a-z-Renyi-MI_free-fermions}
\end{dmath}

Crucially, the correlation matrix $C$ corresponding to the ground state of the model in \cref{eq:Hamiltonian-free-fermions} can be determined analytically~\cite{Bueno2020}:
\begin{equation}
    C_{ij} =
    \begin{cases}
        \frac{(-1)^{i-j}-1}{2\pi\i(i-j)},&i\neq j\\
        \frac{1}{2},&i=j
    \end{cases}.
    \label{eq:a-z-Renyi-MI_free-fermions:Cij}
\end{equation}
This analytical result allows the straightforward construction of $C$ and $C'$ for any partition $(\ell_A,d,\ell_B)$ and facilitates efficient computation of the $\alpha$-$z$ \ac{RMI} using \cref{eq:a-z-Renyi-MI_free-fermions}.
We have performed computations with \textsc{Wolfram Mathematica} using arbitrary precision arithmetic with adaptively chosen precision (see \cref{App:free-fermions:MI-computation} for further details on the numerics).

\subsection{Long-wavelength vs.\ lattice contributions}\label{Sec:free-fermions:lattice-effects}

\begin{figure}[t]
    \centering
    \includegraphics{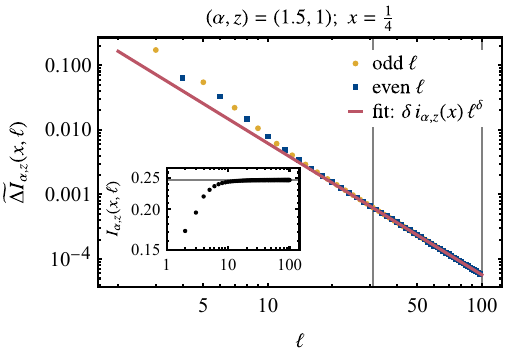}
    \caption{
        Proxy $\widetilde{\Delta I}_{\alpha,z}$ [\cref{eq:lattice-contribution-proxy}] for the lattice contribution $\Delta I_{\alpha,z}$ to the $\alpha$-$z$ Rényi mutual information with $(\alpha,z)=(1.5,1)$ as a function of $\ell=\ell_A=\ell_B$ in a double-logarithmic plot.
        Data points for odd and even $\ell$ are shown as yellow disks and blue squares, respectively.
        The red line represents a linear least-squares fit to the logarithmically rescaled data points with $\ell \in [31,100]$ (gray vertical lines), with slope $\delta = -2.03681(23)$.
        Inset: Convergence of raw data $\Delta I_{\alpha,z}(x,\ell)$ to $I_{\alpha,z}^{\mathrm{CFT}(x)}$ for $\ell\to\infty$ (gray horizontal line).
    }
    \label{fig:proxy-vs-l}
\end{figure}

To systematically analyze lattice effects for different $(\alpha,z)$, we leverage the decomposition of \cref{eq:MI-decomposition}, which attributes any explicit dependence of \ac{MI} on the partition---as opposed to dependence solely on the cross-ratio $x$---to lattice effects.
To simplify the analysis, we focus on the case $\ell_A=\ell_B=\ell$, leaving two relevant length scales: $x$ and $\ell$.
This leaves us with the decomposition
\begin{equation}
    I_{\alpha,z}(x,\ell) = I_{\alpha,z}^{\mathrm{CFT}}(x) + \Delta I_{\alpha,z}(x,\ell).
    \label{eq:azRMI-decomposition}
\end{equation}
However, without direct access to $I_{\alpha,z}^{\mathrm{CFT}}(x)$ and $\Delta I_{\alpha,z}(x,\ell)$ separately, how can these two contributions be separated?

The lattice contributions, $\Delta I_{\alpha,z}$, can be isolated from the computed values of $I_{\alpha,z}$ by noting that any residual dependence on $\ell$ can be attributed to $\Delta I_{\alpha,z}$.
To extract $\Delta I_{\alpha,z}$, we thus compute $I_{\alpha,z}$ at fixed $x$ for various values of $\ell$.
We then numerically evaluate the derivative with respect to $\ell$, which isolates $\partial_\ell\Delta \tilde{I}$.
According to scaling theory, $\Delta I_{\alpha,z}$ is expected to scale as a power-law in $\ell$, $\Delta I_{\alpha,z}\propto\ell^\delta$, where $\delta$ is the scaling exponent.
This motivates defining a proxy for $\Delta I_{\alpha,z}$,
\begin{equation}
    \widetilde{\Delta I}_{\alpha,z}(x,\ell) = \ell\pdv{I_{\alpha,z}}{\ell} = \delta i_{\alpha,z}(x)\ell^\delta + \dotsb,
    \label{eq:lattice-contribution-proxy}
\end{equation}
where the omitted terms represent subleading contributions beyond the dominant power-law.

The scaling exponent $\delta$ directly measures the relevance of $\Delta I_{\alpha,z}$.
If $\delta$ is positive, the lattice contributions grow with $\ell$, while they diminish for $\delta<0$.
Note that $\widetilde{\Delta I}_{\alpha,z}$ can also scale logarithmically, $\propto\log\ell$, such as for $x=1$ in \cref{eq:azRMI_d=0_CFT}.
In such cases, the proxy indicates $\delta = 0$, signaling a breakdown of its utility for following the behavior of $\Delta I_{\alpha,z}$.
However, this breakdown happens in a controlled manner, as it is reliably signaled by a vanishing of $\delta$ and we only need to take care in correctly interpreting $\delta=0$ as indicating logarithmic dependence of the lattice contribution on $\ell$.

Figure~\ref{fig:proxy-vs-l} illustrates how the proxy $\widetilde{\Delta I}_{\alpha,z}(x,\ell)$ can be used to extract the scaling exponent $\delta$.
For sufficiently large $\ell$, $\widetilde{\Delta I}_{\alpha,z}$ exhibits a power-law dependence on $\ell$ for fixed $x$, consistent with the expected scaling behavior.
At smaller $\ell$, however, deviations appear, caused by subleading terms.
As shown in \cref{fig:proxy-vs-l}, $\delta$ is extracted from a linear least-squares fit to the logarithmically rescaled data.
Note that for some $(\alpha,z)$ and $x$, significant odd-even effects in $\ell$ necessitate separate fits for odd and even values of $\ell$.
In the example, we find $\delta = -2.03681(23)$ from a fit to all data points $\ell\geq 31$.

\subsection{Systematic analysis}\label{Sec:free-fermions:systematic-analysis}

\begin{figure}[t]
    \centering
    \subfloat{\label{fig:azRMI-phase-diagram:large_l}}
    \subfloat{\label{fig:azRMI-phase-diagram:small_l}}
    \includegraphics{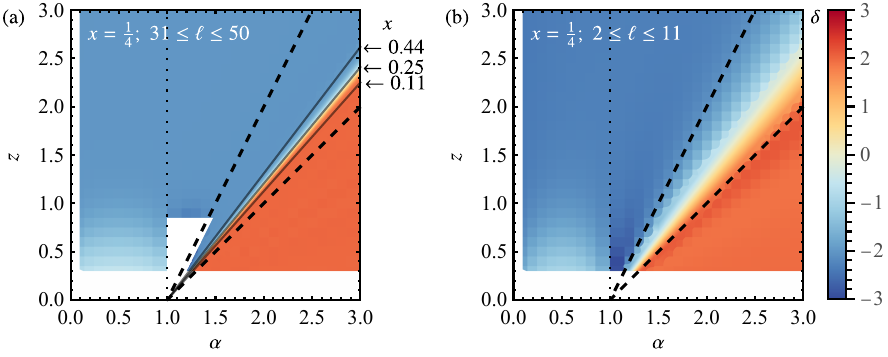}
    \caption{
        \Phasediagrams{} of the relevance of lattice effects in the $\alpha$-$z$ Rényi mutual information of a partition with $x=1/4$ and $\ell_A=\ell_B=\ell$ for the ground state of massless free fermions.
        The relevance is measured by the scaling exponent $\delta$ (see legend to the right) of the lattice contribution $\Delta I_{\alpha,z}$, as defined by the proxy $\widetilde{\Delta I}_{\alpha,z}\sim\ell^\delta$ [\cref{eq:lattice-contribution-proxy}], extracted from fits to the range (a) $31\leq\ell\leq 50$ and (b) $2\leq\ell\leq 11$.
        In (a) solid black lines indicate how the position of the transition from negative (irrelevant) to positive (relevant) scaling exponent $\delta$  changes for choices of $x$ different from $x=\frac{1}{4}$. In both panels the black dashed lines represent the boundaries of the regions I, II and III illustrated in \cref{fig:azRMI-phase-diagram_summary}.
        Note that $\delta=0$, as extracted from the proxy, actually indicates a logarithmic scaling with $\ell$: $\Delta I_{\alpha,z}\sim\log\ell$.
        In (a), a neighborhood of $(1.2,0.6)$ is excluded due to $\widetilde{\Delta I}_{\alpha,z}$ changing sign as a function of $\ell$ near the fitting range before becoming a power law again, preventing an estimation of $\delta$ for this range. Estimating it for smaller $\ell$, as in (b), is possible but we observe that the exponent becomes anomalously small. See \cref{fig:App:DeltaI-vs-l:exclusion} for a discussion and explanation.
    }
    \label{fig:azRMI-phase-diagram}
\end{figure}

\Cref{fig:azRMI-phase-diagram} shows the \phasediagram{} of the relevance of the lattice contributions to the $\alpha$-$z$ \ac{RMI} as a function of $\alpha$ and $z$ for $x=1/4$, as quantified by the scaling exponent $\delta$ extracted from the proxy $\widetilde{\Delta I}_{\alpha,z}$.
The $\alpha$-$z$ plane exhibits two regions separated by a sharp transition: one with $\delta\approx-2$ (blue) and another with $\delta\approx+2$ (red).
In the former region, lattice contributions decay as a power law $\sim\ell^{-2}$, while in the latter, they dominate growing as $\sim\ell^2$ (see \cref{fig:App:DeltaI-vs-l}).
The overall picture is independent of $x$, and only the precise position of the transition line varies.
In \cref{fig:azRMI-phase-diagram:large_l}, the transition lines for $x=1/9$ and $4/9$ are indicated as solid black lines in addition to the one matching the data ($x=1/4$).
The transition lines can be parameterized as
\begin{equation}
    z_c = \eta(x) (\alpha-1),
\end{equation}
where the slope $\eta$ increases with $x$.
For $x=1$, we find numerically that $\eta(1)=2$ in agreement with the analytical and model-independent prediction derived from the divergence of the logarithmic term in \cref{eq:azRMI_d=0_CFT} along $z = 2(\alpha-1)$, see \cref{fig:App:azRMI-phase-diagram:x=1} in the Appendix.
Thus, the transition lines span the region between $\eta(0)=1$ and $\eta(1)=2$, whose boundaries are indicated by black dashed lines in \cref{fig:azRMI-phase-diagram}.

\begin{figure}[t]
    \centering
    \includegraphics{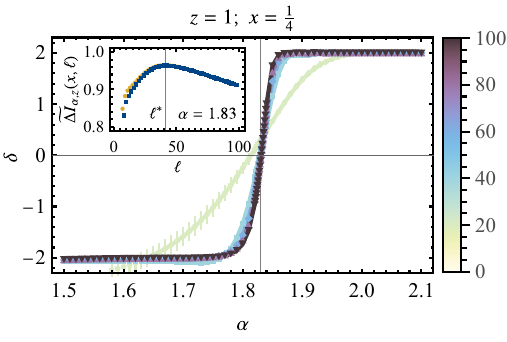}
    \caption{
        Scaling exponent $\delta$ of the lattice contribution extracted form the proxy $\widetilde{\Delta I}_{\alpha,z}\sim\ell^\delta$ [\cref{eq:lattice-contribution-proxy}] to the $\alpha$-$z$ Rényi mutual information near its transition from negative to positive value.
        Exemplary data is shown for $z=1$, $x=1/4$, and $\ell_A=\ell_B=\ell$, but the qualitative behavior is unchanged for other parameter choices.
        The different curves correspond to exponents $\delta$ extracted from different ranges of $\ell$ of size $\Delta\ell = 20$ with the maximal value indicated by the color according to the legend on the right.
        Inset: Proxy as a function of $\ell$ close but to the left of the transition ($\alpha$ marked by a vertical gray line in the main figure).
        As is characteristic of the transition region, the proxy increases at small $\ell$ but then reverses behavior at $\ell=\ell^*$ and decreases with $\ell$.
    }
    \label{fig:delta-vs-alpha}
\end{figure}

This behavior delineates three distinct regions: (I) $\delta<0$ for all $x<1$, where lattice effects act as an irrelevant correction to the long-wavelength behavior and vanish for $\ell\to\infty$; (II) a transition region, where the behavior depends on $x$; and (III) $\delta>0$ for all $x$, where lattice effects are relevant and dominate.
In the transition region (II), $\delta$ changes from negative to positive with the transition line depending on $x$.
At the transition itself, $\delta=0$ extracted from the proxy suggests that \ac{MI} scales as $\log\ell$.
This is consistent with \cref{eq:azRMI_d=0_CFT}, where the divergence of the prefactor of the $\log\ell$ term renders that term dominant.
With this analysis, we have quantitatively established the behavior illustrated in \cref{fig:azRMI-phase-diagram_summary}.

A similar picture emerges when considering only small subsystems, $2\leq\ell\leq 11$, as shown in \cref{fig:azRMI-phase-diagram:small_l}.
While regions I and III remain almost unchanged, there is a visible broadening and small shift of the transition in region II, also seen in \cref{fig:delta-vs-alpha}.
This results from subleading contributions to $\widetilde{\Delta I}_{\alpha,z}(x,\ell)$ [\cref{eq:lattice-contribution-proxy}], as seen in the inset of \cref{fig:delta-vs-alpha}.
The broadening of the transition is accompanied by a local enhancement of $\delta$ and thus of the relevance of the lattice effects, near the boundary between regions II and III.
It is indicated by the slightly darker red close to the line $z=\alpha-1$ signifying a ridge of local maxima of $\delta$.
This demonstrates that while $\delta$ can vary depending on whether it is extracted from small and large $\ell$ (see \cref{fig:proxy-vs-l}), this variation is only significant in the transition region (II) and its neighborhood and does not affect the overall \phasediagram{}.

\Cref{fig:delta-vs-alpha} displays the behavior of the scaling exponent $\delta$ when crossing the transition line at fixed $x<1$.
Specifically, we consider a horizontal cut through \cref{fig:azRMI-phase-diagram} at $z=1$ and plot $\delta$, extracted at various length scales $\ell$ (see legend), as a function of $\alpha$.
We observe that the transition sharpens as $\ell$ increases, confirming the existence of a sharp boundary between the regions with $\delta=-2$ and $\delta=+2$ in the limit $\ell\to\infty$.
Close to but before the transition, the sign of $\delta$ may change depending on the length scale $\ell$ at which it is extracted.
This phenomenon is reflected in the behavior of the proxy $\widetilde{\Delta I}_{\alpha,z}$ as a function of $\ell$, shown in the inset of \cref{fig:delta-vs-alpha}.
At small $\ell$, $\widetilde{\Delta I}_{\alpha,z}$ increases with $\ell$, but eventually turns and decays at larger $\ell$.
This scale-dependent behavior is a defining feature of the transition region and will be made more precise below through the introduction of the \enquote{resolution length scale} associated with each \ac{MI-ad} variant.

\subsection{The resolution length scale}\label{Sec:free-fermions:resolution-length-scale}

\begin{figure}[t]
    \centering
    \includegraphics{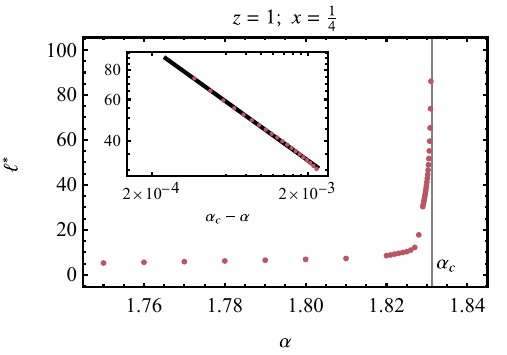}
    \caption{
        Resolution length scale $\ell^*_{\alpha,z}(x)$ as a function of $\alpha$ for $z=1$, $x=1/4$, and $\ell_A=\ell_B$ extracted as demonstrated in the inset of \cref{fig:delta-vs-alpha}.
        It diverges near $\alpha_c\approx1.83128(2)$ (gray vertical line).
        Inset: Double-logarithmic plot of $\ell^*$ as a function of $\alpha_c-\alpha$ together with the least-squares fit.
    }
    \label{fig:length-scale_vs_alpha}
\end{figure}

To explain the presence of two regions with distinct behavior and a transition region, where the behavior depends on both $x$ and $\ell$, we introduce the \emph{resolution length scale} $\ell^*$.
As shown in the inset of \cref{fig:delta-vs-alpha}, $\widetilde{\Delta I}_{\alpha,z}$ reveals two regimes in $\ell$: one where it increases with $\ell$ and another where it decreases.
This behavior naturally suggests a critical length scale $\ell^*$ that separates the two regimes.
For $\ell<\ell^*$, the given variant of \ac{RMI} (at fixed $x$) predominantly reflects lattice effects, whereas for $\ell>\ell^*$, it primarily captures long-wavelength information.
We estimate $\ell^*$ as the position of the maximum of $\widetilde{\Delta I}_{\alpha,z}(\ell)$ which is reliable near but not too close to the transition.
Far from the transition, $\ell^*$ becomes small, and its discrete nature, along with odd/even effects visible in the inset of \cref{fig:delta-vs-alpha} and additional subleading contributions to \ac{MI}, prevents an accurate estimation.
We thus require that $\widetilde{\Delta I}_{\alpha,z}(\ell)$, evaluated separately at odd/even $\ell$, have a single maximum each, with the positions (rounded to the nearest integer) of the two maxima differing by no more than $1$.
Conversely, near $\alpha_c$, $\widetilde{\Delta I}_{\alpha,z}$ becomes extremely flat, similarly again impeding the estimation of $\ell^*$.
As a criterion to determine how near we can go to $\alpha_c$, we here consider the curvature of $\widetilde{\Delta I}_{\alpha,z}$, which we require to be smaller or equal to $-10^{-5}$.

By analyzing the resolution length scale $\ell^*$ for different variants $I_{\alpha,z}(A:B)$ and fixed partitions, we gain insight into how the distinct regions and their behaviors arise.
In the region where $\delta=-2$, $\ell^*$ vanishes because $\widetilde{\Delta I}_{\alpha,z}$ decreases monotonically with $\ell$.
Conversely, in the region where $\delta=+2$, $\ell^*$ diverges, indicating that $\widetilde{\Delta I}_{\alpha,z}$ increases even for arbitrarily small $\ell$ (apart from artifacts at $\ell=1$).
\Cref{fig:length-scale_vs_alpha} shows $\ell^*$ as a function of $\alpha$ close to the transition for $z=1$, $x=1/4$, and $\ell_A=\ell_B$.
Within the transition region, $\ell^*$ is nonzero and, for each $x$, diverges at the specific transition.

The inset of \cref{fig:length-scale_vs_alpha} shows that the divergence is well described by a power law as $\alpha\to\alpha_c$.
To extract the value of $\alpha_c$ where $\ell^*$ diverges, we perform a least-squares fit over the interval $[1.829, 1.8309]$ using the power-law model
\begin{equation}
    \ell^*(\alpha) = \lambda(\alpha_c-\alpha)^{-\nu}
    \label{eq:power-law_lstar-vs-alpha}
\end{equation}
resulting in $\lambda=1.7(1)$, $\alpha_c=1.83128(2)$, and $\nu=0.48(1)$ for a coefficient of determination $R^2=0.999949$.
The fit is shown as a black solid line in the inset.

\section{Transverse-field Ising model}\label{Sec:TFIM}

Although we have argued, based on conformal symmetry and the definition of the $\alpha$-$z$ \ac{RMI}, that the results depicted in \cref{fig:azRMI-phase-diagram_summary} are model-independent, most of our concrete findings on the behavior of the \ac{MI-ad} variants within each region have been obtained from calculations on a free-fermion model.
The purpose of this section is two-fold.
First, we provide evidence that our observations are model-independent and, in particular, also apply to interacting systems, by showing that the same picture emerges for the \ac{TFIM}~\cite{Sachdev2011} at its critical point.
Second, we demonstrate that our observations remain relevant at subsystem sizes where \ac{MI-ad} variants are easily accessible for interacting systems using standard numerical techniques even without specific optimization.

The TFIM describes spin-$1/2$ degrees of freedom on a chain, with Hamiltonian
\begin{equation}
    \Ham_{\mathrm{TFIM}} = -J\sum_{\langle i,j\rangle}s_i^x s_j^x - g\sum_j s_j^z,
    \label{eq:TFIM:Hamiltonian}
\end{equation}
where $s_i^\mu$, $\mu=x,y,z$, are the components of the spin operator $\vec{s}_i$ on site $i$.
We set $J=1$, so that the critical point lies at $g=1$, separating the ferromagnetic ($g<1$) from the paramagnetic ($g>1$) phase~\cite{Sachdev2011}.
At the quantum phase transition ($g=1$), the system is gapless and described at low energies by an Ising \ac{CFT} with central charge $c=1/2$~\cite{DiFrancesco1997}.
In the following, we study the $\alpha$-$z$ \ac{RMI} for the ground state of the TFIM at $J=g=1$.

\begin{figure}[t]
    \centering
    \includegraphics{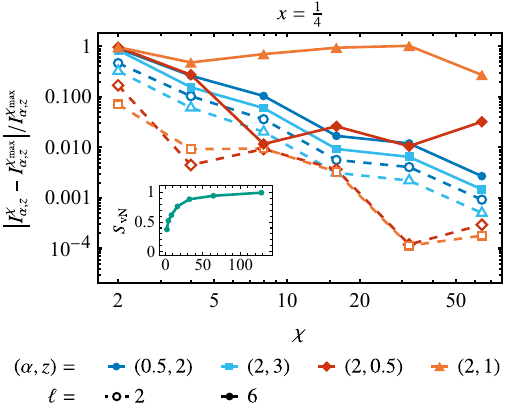}
    \caption{
        Convergence of $\alpha$-$z$ Rényi mutual information, $I_{\alpha,z}^{\chi}(x,\ell)$, for the ground state of the transverse-field Ising model, as a function of bond dimension $\chi$, at $x=1/4$. The relative error is shown as a function of bond dimension on a double-logaritmic plot. Different variants (see legend) are shown for small ($\ell=2$, open markers, dashed lines) and larger ($\ell=6$, filled markers, solid lines) subsystems.
        The inset shows the convergence of the von Neumann entropy for a single site in the two-site unit cell.
    }
    \label{fig:TFIM:convergence}
\end{figure}

The TFIM is, in principle, exactly solvable through a Jordan-Wigner transformation.
However, the nonlocal nature of that duality prevents reducing the \ac{MI} of \emph{disjoint} intervals to single-particle quantities as in \cref{eq:a-z-Renyi-MI_free-fermions}.
Instead the reduced density matrix $\rho_{ACB}$, where $C$ is the interval of length $d$ between $A$ and $B$, would have to be constructed explicitly and the partial trace over $C$ evaluated to obtain $\rho_{AB}$.
This can be avoided by directly working with \acp{MPS}, which at the same time allows us to demonstrate the practical implications of our observations in the setting of a very generic numerical toolbox.
In the following, we thus obtain the ground state of the TFIM using the two-site infinite density matrix renormalization group (iDMRG) algorithm~\cite{White1992,White1993,McCulloch2008,Schollwock2011} implemented in the \textsc{Python} package \textsc{Tenpy}~\cite{Tenpy2024} (see \cref{App:iDMRG} for more details on the numerics), yielding an \ac{iMPS-ad}~\cite{Ostlund1995,Rommer1997,Vidal2007} representation.
We use bond dimensions ranging from $2$ to $2^7=128$ to approximate the gapless ground state, which is sufficient for our purposes because we focus on quantities with support on scales of at most $18$ sites.
The inset of \cref{fig:TFIM:convergence} shows how the von Neumann entropy of one site in the unit cell converges with increasing bond dimension.

Given the \ac{iMPS-ad} representation of the state, the reduced density matrices $\rho_A$, $\rho_B$, and $\rho_{AB}$ can be efficiently computed in their infinite matrix product operator (iMPO) representation (see \cref{App:TFIM:azRMI-iMPO}).
In principle, some specific variants of the $\alpha$-$z$ \ac{RMI} can be computed directly in the iMPO framework without explicitly constructing the reduced density matrices, see \cref{Sec:conclusions}.
However, since our goal is a systematic study of the full two-parameter family, we need to proceed differently.
We construct the reduced density matrices explicitly, enabling us to compute the $\alpha$-$z$ \ac{RMI} for arbitrary $(\alpha,z)$ without additional optimization.
This comes at the cost of limiting the accessible subsystem sizes $\ell_A$ and $\ell_B$ (though not their separation $d$), due to the exponential scaling of computational resources with $\ell_A+\ell_B$.
We have developed the \textsc{Julia} package \textsc{QMICalc}~\cite{QMICalc}, which utilizes \textsc{ITensor}~\cite{Fishman2022,ITensor-r0.3}, to perform these computations for arbitrary \ac{MPS-ad} inputs and \ac{MI-ad} variants.
\Cref{fig:TFIM:convergence} illustrates the convergence of several exemplary \ac{MI-ad} variants with respect to bond dimension.

For small $\ell_A$ and $\ell_B$, the subleading corrections indicated in \cref{eq:lattice-contribution-proxy} are too significant to allow for an accurate extraction of the scaling exponent $\delta$.
However, since we are only interested in whether $\Delta I_{\alpha,z}$ is relevant or irrelevant, we instead consider the ratio between the proxy for lattice contributions, $\widetilde{\Delta I}_{\alpha,z}$, and the full \ac{MI}, $I_{\alpha,z}$:
\begin{equation}
    q(\alpha,z;x,\ell) = \frac{\widetilde{\Delta I}_{\alpha,z}(x,\ell)}{I_{\alpha,z}(x,\ell)}.
    \label{eq:q-definition}
\end{equation}
For irrelevant lattice effects, where $\abs*{\Delta I_{\alpha,z}}\ll \abs*{I^{\mathrm{CFT}}_{\alpha,z}}$, $q$ vanishes for sufficiently large $\ell$, allowing us to identify $q=q_{\mathrm{I}}=0$ with region I.
Conversely, for sufficiently large $\ell$, relevant lattice contributions, $\Delta I_{\alpha,z}\gg I^{\mathrm{CFT}}_{\alpha,z}$, lead to
\begin{equation}
     q(\alpha,z;x,\ell) \approx \frac{\delta i_{\alpha,z}(x)\ell^\delta + \dotsb}{i_{\alpha,z}(x)\ell^\delta + \dotsb} \rightarrow \delta.
\end{equation}
Although $\delta$ is unknown and cannot be reliably determined at small $\ell$, we estimate $q_{\mathrm{III}}$, the value of $q$ in region III, from the saturation value in region III: $q_{\mathrm{III}}\approx q(3,0.4;1/4,4)\approx 0.81$.
At the transition, where $\abs*{\Delta I_{\alpha,z}}\approx \abs*{I^{\mathrm{CFT}}_{\alpha,z}}$, $q$ is expected to either take an intermediate value between $q_{\mathrm{I}}$ and $q_{\mathrm{III}}$ or diverge, depending on the relative sign of the two contributions.

\begin{figure}[t]
    \centering
    \includegraphics{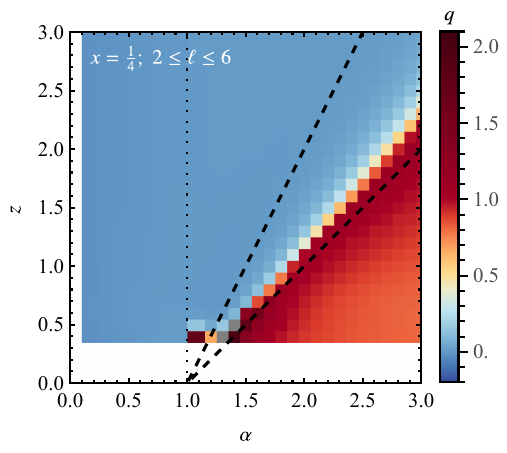}
    \caption{
        \Phasediagram{} of the relevance of lattice contributions to the $\alpha$-$z$ Rényi mutual information of a partition with $x=1/4$ and $\ell=4$ for the ground state of the critical transverse-field Ising model.
        The relevance is quantified by $q(\alpha,z;x,\ell)$ [\cref{eq:q-definition}], where the partial derivative with respect to $\ell$ is approximated by a central difference quotient with support points $\ell=2,6$.
        Black dashed lines indicate the boundaries of the regions defined in \cref{fig:azRMI-phase-diagram_summary}.
        The color scale is aligned with that of \cref{fig:azRMI-phase-diagram}, so that irrelevant ($q=q_{\mathrm{I}}=0$) and relevant ($q=q_{\mathrm{III}}\approx 0.81$) lattice contributions correspond to $\delta=-2$ and $\delta=2$, respectively.
        Extreme values are omitted and shown in gray.
    }
    \label{fig:TFIM:q-vs-a-z}
\end{figure}

\Cref{fig:TFIM:q-vs-a-z} shows $q$ as a function of $\alpha$ and $z$ for $x=1/4$, computed at $\ell=4$.
The color scheme matches that in \cref{fig:azRMI-phase-diagram}, with $q = q_{\mathrm{I}}$ shown in light blue and $q = q_{\mathrm{III}}$ in orange.
We observe that the free-fermion result from \cref{fig:azRMI-phase-diagram:small_l} is qualitatively reproduced.
The ratio $q$ saturates near $q_{\mathrm{I}}=0$ in region I and near $q_{\mathrm{III}}\approx 0.81$ in region III, exhibiting a sharp transition within region II.
\Cref{fig:App:TFIM:q-vs-a-z:further-x} additionally displays analogous results for other choices of $x$, establishing the same qualitative behavior of the transition line within region II as indicated in \cref{fig:azRMI-phase-diagram:large_l}.
Notably, the ridge of maxima near $z=\alpha-1$, i.e., at the boundary between regions II and III, observed for free fermions is also reproduced in the TFIM.
This provides evidence for the model-independence of our main results, as summarized in \cref{fig:azRMI-phase-diagram_summary}.
In contrast, we expect the precise values of $\delta$ and the precise position of the transition line for a given $x$ to depend on the underlying \ac{CFT} as well as the microscopic details of the model.

\section{Discussion and outlook}\label{Sec:conclusions}
\acresetall

The implications of our analysis for applications are most direct in the choice of a \ac{RMI} variant (\cref{fig:summary}).
Variants in region~I that simultaneously satisfy the \ac{DPI} [\cref{eq:DPI}] provide the most reliable access to long-wavelength correlations.
Here, lattice contributions appear only as irrelevant corrections and, if needed, can be removed by extrapolating in subsystem size using the scaling exponent $\delta$ extracted from the proxy introduced in \cref{Sec:free-fermions:lattice-effects}.
When a controlled sensitivity to short-distance features is desirable, the intersection of region~II with the \ac{DPI-ad} region offers a useful dial: the accessible information depends on the partition $(\ell_A,d,\ell_B)$, with the resolution length $\ell^*$ from \cref{Sec:free-fermions:resolution-length-scale} serving as an operational criterion for whether a given setup resolves \ac{CFT-ad} data.
By contrast, region~III has no overlap with the \ac{DPI-ad} region and is dominated by lattice-scale contributions that diverge with $\ell$, rendering it unsuitable both as a faithful correlation measure and as a probe of long-wavelength physics.

Approximations in numerical simulations can introduce cutoffs that manifest as artificial length scales, mimicking lattice effects already at smaller $\ell$.
Through their competition with the subsystem size, they additionally affect the resolution length $\ell^*$. 
Specifically for \acp{MPS}, a finite bond dimension $\chi$ induces an effective correlation length $\xi_\chi\propto\chi^\kappa$~\cite{Tagliacozzo2008,Pollmann2009,Pirvu2012}, which in our simulations of the \ac{TFIM} leads to a variant-dependent convergence of $I_{\alpha,z}$ with~$\chi$ (\cref{fig:TFIM:convergence}). 
This dependence correlates strongly with the region assignment, as corroborated by our analysis in \cref{App:TFIM:convergence-behavior}: variants from region~I converge robustly, those from region~II exhibit scale-dependent behavior, and those from region~III converge poorly and often non-monotonically. 
Comparable artificial length scales can also arise in other approximation schemes, e.g., when an explicit UV cutoff is introduced.
For practical purposes, it is therefore advantageous to focus on region~I measures, already because of their favorable convergence properties.

Within the \ac{MPS-ad} framework, $\alpha$-$z$ \ac{RMI} is, in principle, accessible beyond the small subsystems considered here: while explicit reduced-density-matrix computations suffice for some applications, optimized variational schemes tailored to specific $(\alpha,z)$ offer a more scalable route.
The explicit approach scales exponentially with $\ell_A+\ell_B$ and is thus restricted to moderately small subsystems but already yields useful information for variants deep in region~I, where lattice effects are strongly suppressed. 
More scalable options arise, in particular, for the integer sandwiched \ac{RMI} ($\alpha=z\in\mathbb{Z}$), for which the necessary operator powers can be evaluated fully in the infinite matrix product operator (iMPO) formalism if $(\rho_A\otimes\rho_B)^{(1-\alpha)/z}$ is replaced by a variational iMPO ansatz. 
This construction exploits the variational formulation~\cite{Khatri2024}
\begin{equation}
    I_{\alpha,z}(A:B)=\inf_{\sigma_A,\tau_B} D_{\alpha,z}\left.\left(\rho_{AB}\,\right\|\,\sigma_A\otimes\tau_B\right),
\end{equation}
which suggests a DMRG-style optimization over iMPO parameterizations of $(\sigma_A,\tau_B)$ and avoids explicit non-integer powers of $\rho_A$ and~$\rho_B$.

Beyond tensor-network approaches, quantum mutual information could potentially also be accessed either through classical statistical estimators applied to data from numerical simulations, or through measurement-based protocols that operate directly on a quantum system or simulator. 
In the classical setting, estimator techniques have proven successful, particularly when combined with machine learning~\cite{Poole2019,Belghazi2018,Gokmen2021}, and could be adapted to quantum mutual information. 
Protocols based on randomized measurements, such as classical shadows~\cite{Huang2020}, offer another route. 
The latter method provides estimators for expectation values of arbitrary functions of the density matrix from only polynomially many samples. 
While rigorous performance bounds exist only for linear functions, it has been demonstrated that nonlinear ones can nevertheless be inferred from classical shadows~\cite{Zhang2021,Garcia2021}. 
In both approaches, variants from region~I are natural candidates, since their reduced sensitivity to lattice contributions mitigates systematic errors from limited data or approximations.

Several future directions follow from our analysis. 
Away from criticality, gapped phases with finite correlation length emerge, and we expect \ac{MI-ad} variants in region~I to inherit area-law behavior with rapidly decaying $\Delta I$. 
Our \ac{MPS-ad} results, which effectively realize a finite correlation length, already indicate that the qualitative partition into lattice-irrelevant, transitional, and lattice-dominated regimes can persist beyond strictly critical $(1+1)$D \acp{CFT}, corroborated further by the continuous behavior of mutual information under detuning from criticality. 
It is plausible that this picture extends further, for example to higher dimensions or to gapped phases, albeit with different diagnostics. 
Identifying suitable operational surrogates for the long-wavelength and lattice contributions to mutual information in these cases remains an open task. 
Additionally, expressions analogous to \cref{eq:azRMI_d=0_CFT} can be derived for thermal density matrices, suggesting that some of our results might also extend beyond ground states.

Besides generalizations to other settings, a \emph{rigorous characterization of the resolution scale} would sharpen the practical criteria articulated here and deepen our understanding of the underlying mechanism. 
Finally, clarifying the \emph{relationship between the \phasediagram{}, the \ac{DPI}, and area laws} would connect information-theoretic and physical properties with the observed variation in relevance of lattice effects.
Progress along these directions, together with advances in methods for computing mutual information, is likely to broaden the scope of \ac{MI-ad} as a tool in numerical studies.

\section*{Acknowledgements}
The authors thank Marin Bukov, Anastasiia Skurativska, David Sutter, and Stefan Woerner for valuable discussions.

\paragraph{Author contributions}
M.K.J., T.N., and M.H.F. initiated the project and P.M.L., M.H.F., and T.N. conceptualized it.
P.M.L., supported by M.H.F. and T.N. performed the formal analysis and numerical simulations with the help of M.M.D. for the iDMRG calculations and D.E.G. for the development of QMICalc.jl. 
All authors regularly discussed the results.
P.M.L. wrote the manuscript on which all other authors commented.

\paragraph{Funding information}
P.M.L. acknowledges support by the European Union (ERC, QuSimCtrl, 101113633).
D.E.G was supported by the NSF-Simons National Institute for Theory and Mathematics in Biology (NITMB) Fellowship supported via grants from the NSF (DMS-2235451) and Simons Foundation (MPS-NITMB-00005320).
T.N. acknowledges support from the Swiss National Science Foundation through a Consolidator Grant (iTQC, TMCG-2\_213805) and a Quantum grant (20QU-1\_225225). 

\noindent Views and opinions expressed are however those of the authors only and do not necessarily reflect those of the European Union or the European Research Council Executive Agency. Neither the European Union nor the granting authority can be held responsible for them.

\paragraph{Supplementary Material}
The data presented here and the code for reproducing it are publicly available~\cite{SDC}.

\begin{appendix}
\numberwithin{equation}{section}

\section{Massless free fermions}\label{App:free-fermions}
\subsection{Computing mutual information}\label{App:free-fermions:MI-computation}

To compute the $\alpha$-$z$ \ac{RMI} for free fermions, we have implemented \cref{eq:a-z-Renyi-MI_free-fermions:C,eq:a-z-Renyi-MI_free-fermions:Cprime,eq:a-z-Renyi-MI_free-fermions,eq:a-z-Renyi-MI_free-fermions:Cij} in \textsc{Wolfram Mathematica}.
In \cref{eq:a-z-Renyi-MI_free-fermions}, we replaced $\Tr\log$ by the mathematically equivalent $\log\det$ in order to avoid computing the matrix logarithm.

Because the matrices $\id-C$ and $\id-C'$ are often ill-conditioned, care has to be taken when evaluating \cref{eq:a-z-Renyi-MI_free-fermions}.
To handle those cases reliably, we use arbitrary precision arithmetic with a working precision (number of digits) that we adapt to the choice of $\alpha$ and $z$.
We determine the necessary working precision by computing $I_{\alpha,z}$ at a given $x$ and maximal $\ell$ on a coarse grid in the range of interest $0<\alpha,z\leq 3$ and incrementally increase the working precision from $250$ to $600$ in steps of $50$ and subsequently in steps of $100$ until convergence of the output up to an error of at most $10^{-16}$ is achieved.
The result is then interpolated in $\alpha$ and $z$ and rounded to $50$, giving an upper bound to the working precision for each $\alpha$-$z$ \ac{RMI} required to achieve an output precision of $10^{-16}$.
A minimum of working precision $250$ is used in all cases, even if a lower might have been sufficient for the desired output precision.

\subsection{Relevance of lattice contributions}\label{App:free-fermions:lattice-relevance}

\begin{figure}[t]
    \centering
    \subfloat{\label{fig:App:DeltaI-vs-l:region-I}}
    \subfloat{\label{fig:App:DeltaI-vs-l:region-III}}
    \subfloat{\label{fig:App:DeltaI-vs-l:on-transition}}
    \subfloat{\label{fig:App:DeltaI-vs-l:exclusion}}
    \includegraphics{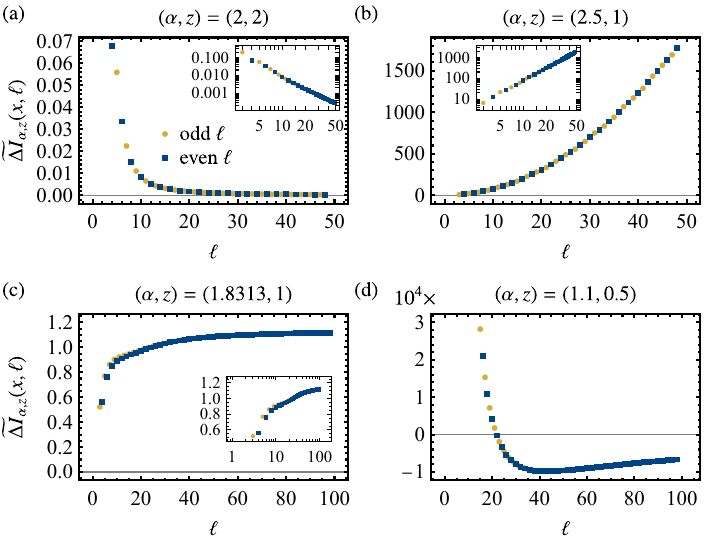}
    \caption{
        Proxy $\widetilde{\Delta I}_{\alpha,z}$ [\cref{eq:lattice-contribution-proxy}] for the lattice contribution $\Delta I_{\alpha,z}$ to the $\alpha$-$z$ Rényi mutual information as a function of $\ell=\ell_A=\ell_B$ for $x=1/4$ and 
        (a) $(\alpha,z)=(2,2)$ in region I, (b) $(\alpha,z)=(2.5,1)$ in region III, (c) $(\alpha,z)=(1.8313,1)$ on the transition line as determined in \cref{Sec:free-fermions:resolution-length-scale}, and (d) $(\alpha,z)=(1.1,0.5)$ in the region excluded in \cref{fig:azRMI-phase-diagram:large_l}.
        Data points for odd and even $\ell$ are shown as yellow disks and blue squares, respectively.
        Insets show the same data on a double-logarithmic scale to emphasize the power-law dependence.
    }
    \label{fig:App:DeltaI-vs-l}
\end{figure}

As discussed in \cref{Sec:free-fermions:lattice-effects}, we quantify the lattice contributions through a proxy quantity $\widetilde{\Delta I}$ [\cref{eq:lattice-contribution-proxy}] which is proportional to the true lattice contribution $\Delta I$ as long as $\Delta I$ is a linear combination of power laws in $\ell$.
This is the case for most $(\alpha,z)$, in particular deep in regions I and III as can be seen in \cref{fig:App:DeltaI-vs-l:region-I,fig:App:DeltaI-vs-l:region-III}, respectively.
However, if $\Delta I$ has a logarithmic dependence on $\ell$, which is the case both at the transition between relevant and irrelevant lattice contributions at fixed $x$, as shown in \cref{fig:App:DeltaI-vs-l:on-transition}, and in all of region I for $x=1$, the proxy $\widetilde{\Delta I}$ will be constant in $\ell$ instead and thus not proportional to $\Delta I$.
As discussed in the main text, this does not prevent us from determining the relevance of the lattice contribution from the proxy---to the contrary, it allows us to treat the logarithmic term $\Delta I$ on the same level as the power-law contributions by converting it to a constant term in the proxy $\widetilde{\Delta I}$.

\begin{figure}[t]
    \centering
    \subfloat{\label{fig:App:azRMI-phase-diagram:x=1/9}}
    \subfloat{\label{fig:App:azRMI-phase-diagram:x=4/9}}
    \subfloat{\label{fig:App:azRMI-phase-diagram:x=1}}
    \includegraphics{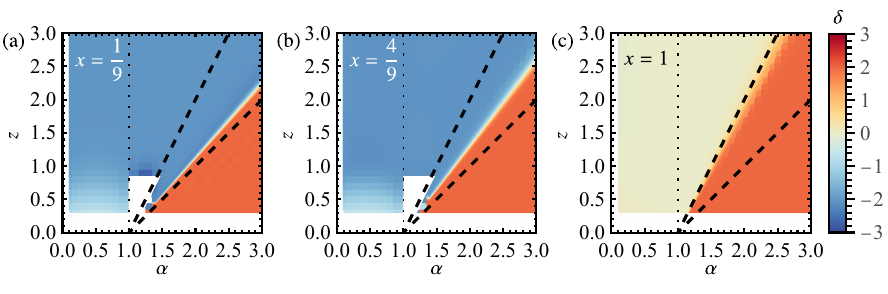}
    \caption{
        \Phasediagrams{} of the relevance of lattice effects in the $\alpha$-$z$ Rényi mutual information of a partition with $\ell_A=\ell_B=\ell$ (a) $x=1/4$, (b) $x=4/9$, (c) $x=1$ ($d=0$) for the ground state of massless free fermions.
        The relevance is measured by the scaling exponent $\delta$ of (a,b) the lattice contribution $\Delta I_{\alpha,z}$ and (c) of the total mutual information, in both cases determined from fits in the range $31\leq\ell\leq 50$.
        See caption of \cref{fig:azRMI-phase-diagram} for more details.
        Note that a vanishing $\delta$ indicates a logarithmic scaling with $\ell$, which is the expected behavior of the universal long-wavelength contribution for $x=1$ due to regularization [see \cref{eq:azRMI_d=0_CFT}].
    }
    \label{fig:App:azRMI-phase-diagram:further-x}
\end{figure}

The leading contribution to $\widetilde{\Delta I}$ can thus be fit by a single power law $\propto\ell^\delta$, where the exponent $\delta$ directly corresponds to the scaling exponent of $\Delta I$, except if it vanishes in which case it indicates a logarithmic dependence of $\Delta I$.
As can be seen in \cref{fig:App:DeltaI-vs-l:region-I,fig:App:DeltaI-vs-l:region-III}, the coefficient in the power law typically has the same sign as the exponent leading to a positive $\widetilde{\Delta I}$.
In \cref{fig:azRMI-phase-diagram:large_l,fig:delta-vs-alpha}, we recognize that indeed $\delta=0$ at the transition implying a logarithmic dependence, as discussed above.
\Cref{fig:App:azRMI-phase-diagram:further-x} shows the same \phasediagram{} for different choices of $x$, cf. the solid black lines in \cref{fig:azRMI-phase-diagram:large_l} that are extracted from the data shown in \cref{fig:App:azRMI-phase-diagram:x=1/9,fig:App:azRMI-phase-diagram:x=4/9}.
In particular, \cref{fig:App:azRMI-phase-diagram:x=1} shows $\delta$ for $x=1$.
According to \cref{eq:azRMI_d=0_CFT}, we expect all of region I to depend logarithmically on $\ell$, which matches with the observation that $\delta$, as extracted from $\widetilde{\Delta I}$, vanishes in all of region I.

A practical complication for extracting $\delta$ occurs in the vicinity of $(\alpha,z)=(1.1,0.5)$, see white excluded region in \cref{fig:azRMI-phase-diagram:large_l,fig:App:azRMI-phase-diagram:x=1/9,fig:App:azRMI-phase-diagram:x=4/9}.
In this neighborhood, the $\alpha$-$z$ \ac{RMI} shows a strongly nonmonotonic dependence on $\ell$ with a rapid increase at small $\ell$ and a slower decrease at larger $\ell$ manifesting as a sign change in $\widetilde{\Delta I}$ as depicted in \cref{fig:App:DeltaI-vs-l:exclusion}.
This can be explained by a dependence of the form
\begin{equation}
    I(\ell) = a\ell^\delta + b\ell^\gamma,
\end{equation}
with $\delta,\gamma<0$, $a>0$, and $b<0$ if the subleading term is comparable in magnitude to the leading term for $\ell$ on the order of $10$ or larger ($\sim20$ in the specific example), i.e., if $\abs*{b}$ is at least an order of magnitude larger than $\abs*{a}$.
The latter implies that $\delta$ cannot be extracted from the chosen fitting range ($31\leq\ell\leq 50$), which contains the minimum of $\widetilde{\Delta I}$ almost perfectly in its center.
Even values of $\ell$ as displayed in \cref{fig:App:DeltaI-vs-l:exclusion} would not be sufficient to eliminate the effect of the subleading term (we find $\delta\approx -0.7$ when fitting the data in the range $81\leq\ell\leq 100$ for the example where this data is available).
Note that extracting $\gamma$ at small $\ell$ works a bit more reliably and we observe that in the corresponding region in \cref{fig:azRMI-phase-diagram:small_l}, the exponent $\sim -3$ is actually an estimate of the exponent $\gamma$ and not $\delta$.

\section{The transverse-field Ising model}\label{App:TFIM}

\subsection{Infinite density matrix renormalization group}\label{App:iDMRG}

Infinite density matrix renormalization group calculations are performed using \textsc{Tenpy}~\cite{Tenpy2024} (version v0.9.0) with a two-site unit cell and a ten-site environment.
We use a mixer of amplitude $10^{-3}$, decay length $5$ for $50$ steps, and a singular-value-decomposition cutoff of $10^{-10}$ ($10^{-12}$ for the largest bond dimension $128$).
The bond dimension is increased incrementally in powers of $2$, ranging from $2$ to $128$.

\subsection{Computing mutual information using infinite matrix product states}\label{App:TFIM:azRMI-iMPO}

We consider a translationally invariant \ac{MPS} of a \ac{1D-ad} chain of size $N$ with a single effective $D$-dimensional degree of freedom $i_n$ per site $n$ (in our case, each site corresponds to an instance of the unit cell made up of two adjacent spin-1/2):
\begin{equation}
    \ket{\Psi} = \sum_{[i]}\Tr\left(\prod_{n=1}^N A_{i_n}\right)\ket{[i]_N},
\end{equation}
where $[i]_N$ denotes the collection of physical indices $i_n=1,2,\dotsc,D$ with $n=1,2,\dotsc,N$, and $A_{i_n}$ are $\chi\times\chi$ matrices characterizing the translation invariant state.
The full density matrix $\rho$ is then given by the \ac{MPDO}
\begin{equation}
    \rho = \sum_{[i]_N,[j]_N}\Tr\left(\prod_{n=1}^N M_{i_n,j_n}\right)\op{[i]_N}{[j]_N},
\end{equation}
where
\begin{equation}
    M_{ij} = A_i^{\phantom{*}}\otimes A_j^*.
\end{equation}

Due to translation invariance, computing the partial trace over site $n$ amounts to replacing the corresponding matrix $M_{i_n,j_n}$ in the product within the trace by the transfer matrix
\begin{equation}
    T = \sum_{i=1}^D M_{i,i}.
\end{equation}
Then the reduced density matrix of a subsystem $A$ of size $\ell_A=N_A$ becomes
\begin{equation}
    \rho_A = \sum_{[i]_A,[j]_A}\Tr\left(M^{(A)}_{[i],[j]}T^{N-N_A}\right)\op{[i]_A}{[j]_A},
\end{equation}
where $[i]_A$ denotes the restriction of physical indices to subsystem $A$ and we abbreviated
\begin{equation}
    M^{(A)}_{[i],[j]} = \prod_{n=1}^{N_A}M_{i_n,j_n}.
\end{equation}

The transfer matrix $T$ has the eigendecomposition
\begin{equation}
    T = \sum_{a=1}^{\chi^2}\lambda_a\op{\lambda_a^r}{\lambda_a^l},\quad 1=\lambda_1>\lambda_2\geq\dotsb\geq\lambda_{\chi^2},
\end{equation}
where we assume that there is only a single eigenvalue with magnitude 1, and that left and right eigenvector are chosen such that $\ip*{\lambda_a^l}{\lambda_b^r}=\delta_{ab}$.
As a consequence,
\begin{equation}
    \lim_{N\to\infty} T^N = \sum_{a=1}^{\chi^2}\lambda_a^N\op{\lambda_a^r}{\lambda_a^l} = \op{\lambda_1^r}{\lambda_1^l},
\end{equation}
such that
\begin{equation}
    \lim_{N\to\infty} \mel{[i]_A}{\rho_A}{[j]_A} = \mel{\lambda_1^l}{M^{(A)}_{[i],[j]}}{\lambda_1^r}.
\end{equation}
This gives us immediate access to the \emph{matrix elements} of the reduced density matrix $\rho_A$ given the \ac{iMPS} state as encoded in the matrices $A_i$.

\begin{figure*}[t]
    \centering
    \subfloat{\label{fig:App:TFIM:q-vs-a-z::x=1/9}}
    \subfloat{\label{fig:App:TFIM:q-vs-a-z::x=1}}
    \includegraphics{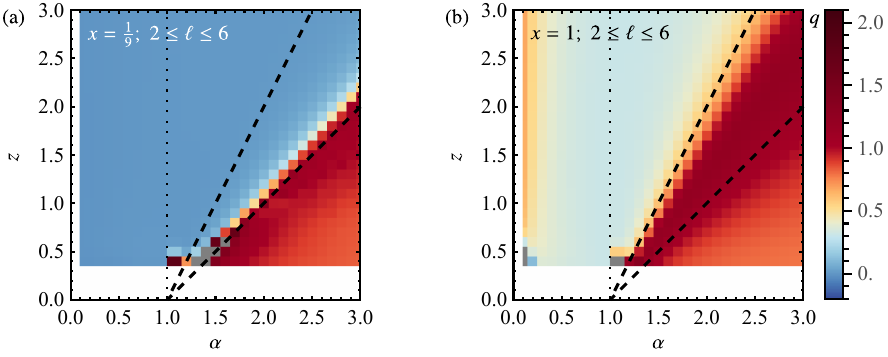}
    \caption{
        (a) \Phasediagram{} of the relevance of lattice contributions to the $\alpha$-$z$ Rényi mutual information of a partition with $x=1/9$ and $\ell=4$ for the ground state of the the critical transverse-field Ising model.
        The relevance is quantified by $q(\alpha,z;x,\ell)$.
        See caption of \cref{fig:TFIM:q-vs-a-z} for more details.
        (b) \Phasediagram{} of $q(\alpha,z;x,\ell)$ for the total mutual information (MI) in the special case $x=1$ ($d=0$), where the universal long-wavelength contribution is influenced by lattice effects through regularization.
        For adjacent subsystems, the proxy $\widetilde{\Delta I}_{\alpha,z}$ captures the total MI  but we can still distinguish the $\propto\log\ell$ behavior [\cref{eq:azRMI_d=0_CFT}] in region I leading to $q\lesssim 0.5$ from the diverging MI with $q\gtrsim1$ where nonuniversal terms take over.
    }
    \label{fig:App:TFIM:q-vs-a-z:further-x}
\end{figure*}

Similarly, we find the matrix elements of the reduced density matrix of the disjoint subsystem consisting of part $A$ of size $\ell_A=N_A$ and part $B$ of size $\ell_B=N_B$ separated by $d$ sites to be
\begin{equation}
    \lim_{N\to\infty} \mel{[i]_{AB}}{\rho_{AB}}{[j]_{AB}} =
    \sum_{a=1}^{\chi^2}\lambda_a^L\mel{a_1^l}{M^{(A)}_{[i],[j]}}{\lambda_a^r}\mel{\lambda_a^l}{M^{(B)}_{[i],[j]}}{a_1^r}.
\end{equation}
Since the $\alpha$-$z$ \ac{RMI} for a given partition only depends on $\rho_A$, $\rho_B$, and $\rho_{AB}$, this provides all the necessary information to compute it from the \ac{iMPS}.
In practice, we only need to compute the matrices $M_{ij}$ and the transfer matrix and its eigendecomposition once per state, i.e., once for a given bond dimension $\chi$, and $M^{(A)}$, $M^{(B)}$ once per partition.
Subsequently, either the the matrix elements of the reduced density matrices explicitly or the action of the reduced density matrix on a vector can be computed efficiently within the \ac{MPS} framework.
This functionality, together with convenient functions and types for setting up the states and partitions, is implemented in the \textsc{Julia} package \textsc{QMICalc}~\cite{QMICalc} which utilizes the \textsc{ITensor}~\cite{Fishman2022,ITensor-r0.3} (version v0.8.0) package to perform the underlying tensor-network calculations.

\subsection{Relevance of lattice contributions}

In addition to \cref{fig:TFIM:q-vs-a-z}, which shows the \phasediagram{} of the relevance of lattice contributions in the TFIM for partitions with $x=1/4$ as deduced from $q$ defined in \cref{eq:q-definition}, we show the analogous data for $x=1/9$ and the special case $x=1$ in \cref{fig:App:TFIM:q-vs-a-z::x=1/9,fig:App:TFIM:q-vs-a-z::x=1}, respectively.
The transition, indicated by a change of color from blue(ish) to red, clearly shifts toward the lower black dashed line analogously to the observation made for massless free fermions in \cref{fig:azRMI-phase-diagram:large_l,fig:App:azRMI-phase-diagram:further-x}.

For $x=1$, the quantity $q$ is expected to behave as
\begin{equation}
    q(\alpha, z; x=1, \ell) \sim \frac{1}{\log\ell},
\end{equation}
which vanishes for $\ell\to\infty$ but much more slowly than for $x<1$.
This manifests as the larger value of $q$ in region I in \cref{fig:App:TFIM:q-vs-a-z::x=1}.
However, it is still significantly smaller than the saturation value in region III, $q_{\mathrm{III}}\approx 0.81$.

\subsection{Convergence behavior of mutual information}\label{App:TFIM:convergence-behavior}

As discussed in \cref{Sec:conclusions}, truncating an \ac{iMPS} describing the ground state at a critical point at finite bond dimension $\chi$ introduces an effective (finite) correlation length $\xi_\chi\propto\chi^\kappa$.
Similarly to how the presence of the lattice breaks conformal invariance and introduces contributions to the \ac{MI} that go beyond the universal \ac{CFT} value, we expect that the finite bond dimension leads to finite-correlation-length contributions in the form
\begin{equation}
    I^\chi = I^{\mathrm{lat}} + \Delta I(\xi_\chi),
\end{equation}
analogous to \cref{eq:MI-decomposition}.
As a crude approximation, we can take $I^{\mathrm{lat}} \approx I^{\chi_{\mathrm{max}}}$, where $\chi_{\mathrm{max}}=128$ is the largest bond dimension used in the simulations.
\Cref{fig:TFIM:convergence} demonstrates that the resulting quantity follows a power-law behavior as a function of $\chi$ and thus $\xi_\chi\propto\chi^\kappa$, i.e.,
\begin{equation}
    \Delta I(\xi_\chi) \approx \abs{I^\chi - I^{\chi_{\mathrm{max}}}}\propto \xi_\chi^{\delta_\xi}.
\end{equation}

\begin{figure}[t]
    \centering
    \includegraphics{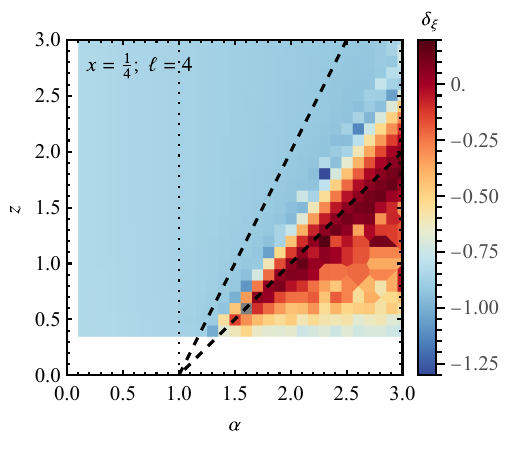}
    \caption{
        \Phasediagram{} of the relevance of the finite-bond corrections to the $\alpha$-$z$ Rényi mutual information, for a partition with $x=1/4$ and $\ell=4$ in the ground state of the critical transverse-field Ising model. The data were obtained using density-matrix-renormalization-group calculations.
        The relevance is measured through the scaling exponent $\delta_\xi$ of the corrections due to the finite bond dimension.
        A larger magnitude of $\delta_\xi$ indicates lower sensitivity to truncation at finite bond dimension and thus faster convergence.
    }
    \label{fig:App:TFIM:convergence-vs-a-z}
\end{figure}

\Cref{fig:App:TFIM:convergence-vs-a-z} shows the exponent $\delta_\xi$ for different variants of $\alpha$-$z$ \ac{RMI}, resulting in a \phasediagram{} of the relevance of \emph{corrections due to the finite bond dimension}.
The exponent is obtained by fitting $\log\abs*{I_{\alpha,z}^\chi - I_{\alpha,z}^{\chi_{\mathrm{max}}}}$ of a partition with $x=1/4$ and $\ell=4$ for the ground state of the \ac{TFIM} as a function of $\log\xi_\chi$ obtained directly from the iDMRG calculation with a linear model.
The \phasediagram{} can thus be interpreted in two ways: it characterizes how sensitive the different $\alpha$-$z$ \ac{RMI-ad} variants are to \ac{MPS-ad} approximations controlled by the bond dimension, which at the same time determines how quickly a particular variant converges when the underlying \ac{MPS} is obtained through density-matrix-renormalization-group calculations.

\end{appendix}



\end{document}